\documentclass[preprint]{ptephy_v1}
\usepackage{graphicx}
\usepackage{color,amsmath,multirow,url}

\preprintnumber{J-PARC-TH-0170}

\begin{document}
\title{
  Classifying Topological Charge in SU(3) Yang-Mills Theory with Machine Learning}

\author[1]{Takuya Matsumoto}
\affil{Department of Physics, Osaka University, Toyonaka, Osaka 560-0043, Japan}

\author[1,2]{Masakiyo Kitazawa\thanks{kitazawa@phys.sci.osaka-u.ac.jp}}
\affil{J-PARC Branch, KEK Theory Center, Institute of Particle and Nuclear Studies, KEK, 203-1, Shirakata, Tokai, Ibaraki, 319-1106, Japan}

\author[3]{Yasuhiro Kohno}
\affil{Research Center for Nuclear Physics, Osaka University, Ibaraki, Osaka 567-0047, Japan}

\date{\today}

\begin{abstract}
  We apply a machine learning technique for identifying the
  topological charge of quantum gauge configurations
  in four-dimensional SU(3) Yang-Mills theory.
  The topological charge density measured
  on the original and smoothed gauge configurations with and without
  dimensional reduction is used as inputs for the neural networks (NN)
  with and without convolutional layers.
  The gradient flow is used for the smoothing of the gauge field.
  We find that the topological charge determined at a large flow time
  can be predicted with high accuracy from the data at small flow times
  by the trained NN; for example,
  the accuracy exceeds $99\%$ with the data at $t/a^2\le0.3$.
  High robustness against the change of simulation parameters
  is also confirmed with a fixed physical volume.
  We find that the best performance is obtained when the 
  spatial coordinates of the topological charge density are fully integrated
  out in preprocessing, which implies that our convolutional NN
  does not find characteristic structures in multi-dimensional space
  relevant for the determination of the topological charge.
\end{abstract}

\maketitle

\section{Introduction}
\label{sec:intro}

Quantum chromodynamics (QCD) and other Yang-Mills gauge theories
in four spacetime dimensions can have topologically non-trivial
gauge configurations classified by the topological charge ${\cal Q}$
taking integer values.
The existence of non-trivial topology in QCD is responsible
for various non-perturbative aspects of this theory,
such as the U(1) problem~\cite{Weinberg:1996kr}.
The susceptibility of ${\cal Q}$ also provides an essential
parameter relevant to the cosmic abundance of the axion dark
matter~\cite{Peccei:1977,PRESKILL1983127,ABBOTT1983133}.

The topological property of QCD and Yang-Mills theories has been
studied by numerical simulations of lattice gauge
theory~\cite{Berkowitz:2015aua,Kitano:2015fla,Ce:2015qha,
  Bonati:2015vqz,Petreczky:2016vrs,Frison:2016vuc,Borsanyi:2016ksw,
  Taniguchi:2016tjc,Aoki:2017paw,Alexandrou:2017hqw,Burger:2018fvb,
  Jahn:2018dke,Bonati:2018blm,Giusti:2018cmp}.
Because of the discretization of spacetime, 
gauge configurations on the lattice are, strictly speaking,
topologically trivial.
However, it is known that well-separated topological sectors
emerge when the continuum limit is approached~\cite{Luscher:1981zq}.
Various methods for the measurement of ${\cal Q}$ for gauge
configurations on the lattice have been proposed, which 
are roughly classified into the fermionic and gluonic ones.
In the fermionic definitions the topological charge is defined
through the Atiyah-Singer index theorem~\cite{Atiyah71},
while the gluonic definitions make use of the topological charge
measured on a smoothed gauge field~\cite{Iwasaki:1983bv,Teper:1985rb}.
The values of ${\cal Q}$ measured by various methods
show approximate agreement~\cite{Alexandrou:2017hqw}, which 
indicates the existence of well-separated topological sectors.
In lattice simulations, 
the measurement of the topological charge is also important 
for monitoring topological
freezing~\cite{DelDebbio:2002xa,Aoki:2007ka,Schaefer:2010hu,Luscher:2011kk}.

In the present study, we apply the machine learning (ML) techniques
for analyzing ${\cal Q}$ for gauge configurations on the lattice.
The ML has been applied for various problems in computer science
quite successfully, such as the image recognition, object detection, and
natural language processing~\cite{726791,Krizhevsky:2012,ICML2012Le_73,
  2013arXiv1312.4400L,Szegedy_2015_CVPR,Simonyan15,He_2016_CVPR,
  Girshick_2014_CVPR,Girshick_2015_ICCV,liu2016ssd,
  NIPS2015_5638,Redmon_2016_CVPR,NIPS2013_5021,NIPS2017_7181,
  devlin2018bert,45774,mnih2015humanlevel,silver2017mastering}.
Recently, this technique has also been applied to problems in physics~%
\cite{712178,Baldi:2014kfa,deOliveira:2015xxd,doi:10.7566/JPSJ.85.123706,
  Barnard:2016qma,Tanaka:2016rtu,Carrasquilla2017,Wetzel:2017ooo,
  Mori:2017pne,raissi2017physicsI,
  Huang:2018fzn,Shanahan:2018vcv,Hashimoto:2018ftp,Kashiwa:2018jdi,
  Steinheimer:2019iso,Fukushima:2019qpv}.
In the present study, we generate data by the
numerical simulation of SU(3) Yang-Mills theory 
in four spacetime dimensions, and feed them into
the neural networks (NN).
The NN are trained to predict the value of ${\cal Q}$
by the supervised learning.
We use the convolutional NN (CNN) as well as the simple fully-connected
NN (FNN) depending on the type of the input data.

The first aim of this study is development of an efficient algorithm
for the analysis of ${\cal Q}$ with the aid of the ML.
The second, and more interesting, purpose is the search for 
characteristic local structures
in the four-dimensional space related to ${\cal Q}$ by the CNN.
The CNN is a class of NN that was developed for 
image recognition~\cite{Krizhevsky:2012,Szegedy_2015_CVPR,Simonyan15,He_2016_CVPR}.
The CNN has so-called convolutional layers which
are composed of filters analyzing a small spatial region of
the output of the previous layer.
If features of the data related to the answer
are embedded in the filter window, the convolutional layers can be
trained to output a signal related to these features.
This design of the CNN had turned out to be quite effective
in image recognition, i.e. the analysis of two-dimensional
data~\cite{Krizhevsky:2012,Szegedy_2015_CVPR,Simonyan15,He_2016_CVPR}.
It is thus expected that the CNN is also suitable for the analysis of
four-dimensional data.
It is known that Yang-Mills theories have classical gauge configurations
called instantons, which carry a nonzero topological charge and 
have a localized structure~\cite{Weinberg:1996kr}.
If the topological charge of the quantum gauge configurations is also
carried by instanton-like local objects in four-dimensional space,
a four-dimensional CNN would recognize and make use of them
for the prediction of ${\cal Q}$.
Such an analysis of four-dimensional quantum fields
by ML will open up a new application of this technique.

In this study, we use the topological charge density measured
on the original and smoothed gauge configurations as inputs to the NN.
The smoothing is performed by the gradient flow%
~\cite{Narayanan:2006rf,Luscher:2010iy,Luscher:2011bx}.
We also try dimensional reduction to various dimensions
as preprocessing of the data before feeding them into a CNN or FNN.
For the definition of ${\cal Q}$, we use a gluonic one 
through the gradient flow~\cite{Luscher:2010iy,Luscher:2011bx}.
We find that the NN can estimate the value of ${\cal Q}$
determined at a large flow time with high accuracy
from the data obtained at small flow times.
In particular, we show that a high accuracy is obtained
by multi-channel analysis of the data at different flow times.
We argue that this method can be used for the analysis of ${\cal Q}$
in SU(3) Yang-Mills theory with a reduction in numerical cost.

To evaluate the performance of the NN searching for
high-dimensional data, we compare the resulting accuracies
obtained with and without the dimensional reduction.
From this comparison
we find that the accuracy does not have a
statistically significant dependence on the dimension of the
input data. 
This result implies that the CNN fails in finding characteristic
features related to the topology in multi-dimensional space,
i.e. the quantum gauge configurations do not have such features,
or their signals are too complicated or too weak to be detected
by the CNN.

\section{Organization of this paper}
\label{sec:organization}

In this study, we perform various analyses of the topological charge
${\cal Q}$ by CNN or FNN.
One of them is the analysis of 
the topological charge density $q_t(x)$
in the four-dimensional ($d=4$) space at a flow time $t$
(the definitions of $q_t(x)$ and $t$ will be given in Sec.~\ref{sec:Q}).
We also perform the dimensional reduction of the input data
as preprocessing and analyze them using the NN.
The dimension of the data is reduced to $d=0-3$ by integrating out
spatial coordinates.
For the analysis at $d=0$ we adopt an FNN,
while the data at $d\ge1$ are analyzed by a CNN.
We then compare the accuracy of the trained NN obtained for each input
dimension.
If features in the four-dimensional space are 
too abstract to be recognized by the NN, the dimensional reduction
will improve the accuracy of the trained NN.
On the other hand, if characteristic features are lost by the
dimensional reduction, this procedure would lower the accuracy.

In the present study
we find that the resulting accuracy of the NN is insensitive to
the value of $d$.
Because the numerical cost for the supervised learning is suppressed 
as $d$ becomes smaller, this means that the analysis of the 
$d=0$ data is most efficient.
In this paper, therefore, we first report this most successful
result among various analyses with the ML technique in Sec.~\ref{sec:0}.
The analyses of the multi-dimensional data will then be discussed
later in Secs.~\ref{sec:q(x)} and \ref{sec:reduction}.

Before introducing the ML technique,
we consider simple models to make an estimate of ${\cal Q}$ without ML
in Sec.~\ref{sec:bench}.
These models are used for benchmarks of the trained NN
in Secs.~\ref{sec:0}--\ref{sec:reduction}.

The whole structure of this paper is summarized as follows.
In the next section, we describe the setup of the lattice numerical simulations.
In Sec.~\ref{sec:Q}, we give a brief review of the analysis
of the topology with the gradient flow.
The benchmark models for the classification of ${\cal Q}$
without using ML are discussed in Sec.~\ref{sec:bench}.
The application of ML is then discussed
in Secs.~\ref{sec:0}--\ref{sec:reduction}.
We first consider the analysis of the $d=0$ data 
by FNN in Sec.~\ref{sec:0}.
We discuss the analysis of the four-dimensional field $q_t(x)$
by CNN in Sec.~\ref{sec:q(x)}.
In Sec.~\ref{sec:reduction}, we extend the analysis to 
$d=1,2,3$.
The last section is devoted to discussions.

\section{Lattice setup}
\label{sec:setup}

\begin{table}
  \centering
  \begin{tabular}{ccc}
    \hline\hline
    $\beta$ & $N^4$ & $N_{\rm conf}$ \\ 
    \hline
    6.2     & $16^4$ & 20,000 \\ 
    6.5     & $24^4$ & 20,000 \\ 
    \hline\hline
  \end{tabular}
  \caption{
    Simulation parameters on the lattice:
    the inverse bare coupling $\beta$, the lattice size $N^4$, and
    the number of configurations $N_{\rm conf}$.
  }
  \label{table:setting}
\end{table}

Throughout this paper, we consider 
SU(3) Yang-Mills theory in four-dimensional Euclidean space
with periodic boundary conditions for all directions.
The standard Wilson gauge action is used for generating the gauge
configurations.
We perform the numerical analyses 
at two inverse bare couplings $\beta=6/g^2=6.2$ and $6.5$ with 
lattice volumes $16^4$ and $24^4$, respectively,
as shown in Table~\ref{table:setting}.
These lattice parameters are chosen so that the lattice volumes
in physical units $L^4$ are almost the same on these lattices;
the lattice spacing determined in Ref.~\cite{Giusti:2018cmp}
shows that the difference in the lattice size $L$ is less than $2\%$.
The lattice size $L$ is related to the critical temperature
of the deconfinement phase transition $T_c$ as 
$1/L\simeq0.63T_c$~\cite{Kitazawa:2016dsl}.

We generate $20,000$ gauge configurations for each $\beta$,
which are separated by $100$ Monte Carlo sweeps from each other, 
where one sweep consists of one pseudo-heat bath and
five over-relaxation updates.
For the discretized definition of the topological charge density on
the lattice, we use the operator constructed from the clover
representation of the field strength.
The gradient flow is used for the smoothing of the gauge field.

To estimate the statistical error of an observable on the lattice, 
we use jackknife analysis with the binsize $100$.
We have numerically checked that the auto-correlation length of the
topological charge is about $100$ and $1900$ sweeps for 
$\beta=6.2$ and $6.5$, respectively.
The binsize of the jackknife analysis including $100\times100$
sweeps is sufficiently larger than the auto-correlation length.

\section{Topological charge}
\label{sec:Q}

In the continuous Yang-Mills theory
in four-dimensional Euclidean space,
the topological charge is defined by
\begin{align}
  {\cal Q} &= \int_V d^4x\,q(x),
  \label{eq:Q}
  \\
  q(x) &= -\frac{1}{32\pi^2}\epsilon_{\mu\nu\rho\sigma}
  {\rm tr}\left[F_{\mu\nu}(x)F_{\rho\sigma}(x)\right],
  \label{eq:q(x)}
\end{align}
where $V$ is the four-volume and
$F_{\mu\nu}(x)=\partial_\mu A_\nu(x)-\partial_\nu A_\mu(x)+[A_\mu(x),A_\nu(x)]$
is the field strength.
$q(x)$ is called the topological-charge density with the coordinate $x$
in Euclidean space.

In lattice gauge theory, Eq.~(\ref{eq:Q}) calculated on a gauge
configuration with a discretized definition of Eq.~(\ref{eq:q(x)})
is not given by an integer, but distributes continuously.
To obtain discretized values, one may apply a smoothing
of the gauge field before the measurement of $q(x)$.

\begin{figure}[t]
  \centering
  \vspace{5mm}
  \includegraphics[width=0.49\textwidth,clip]{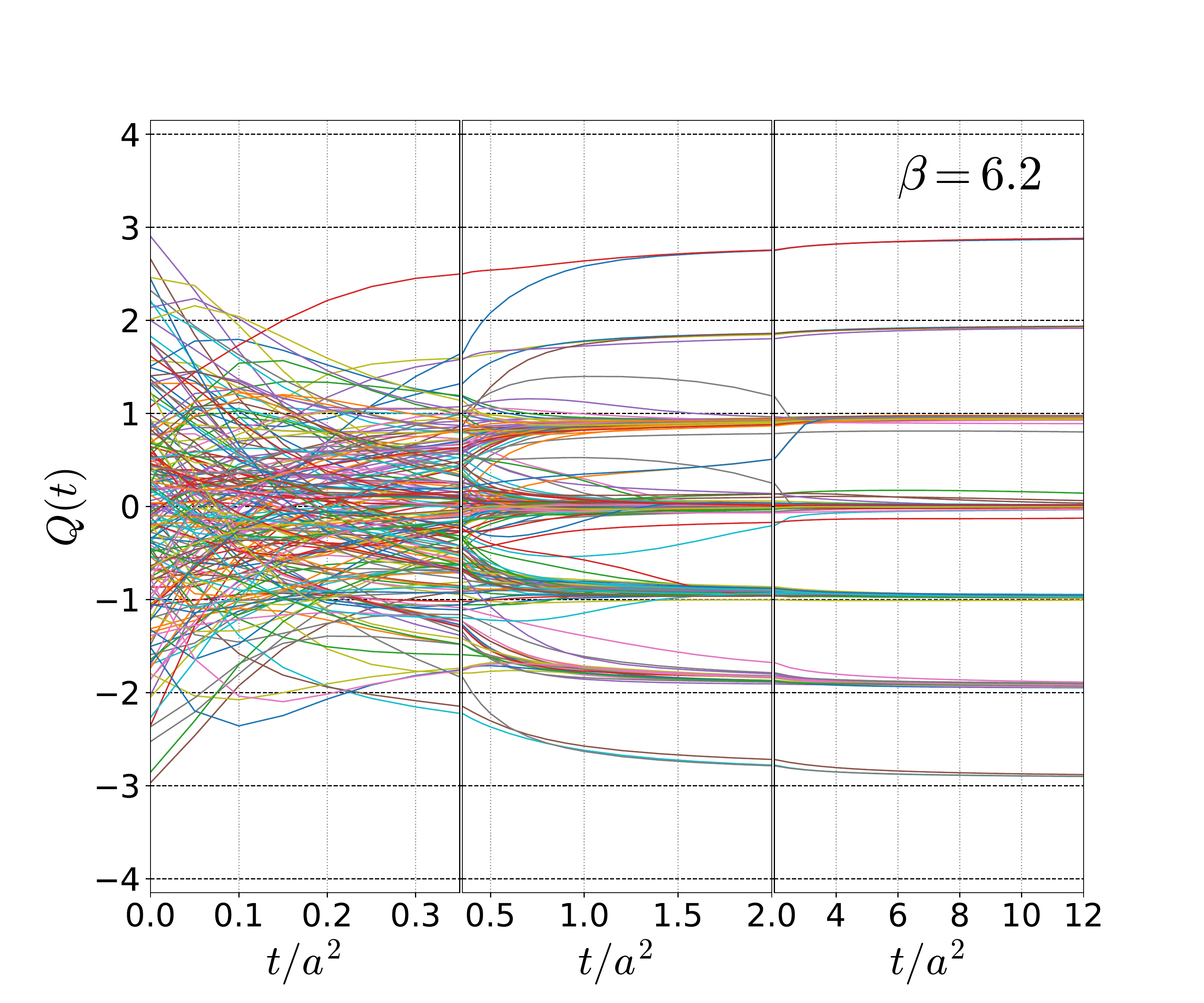}
  \includegraphics[width=0.49\textwidth,clip]{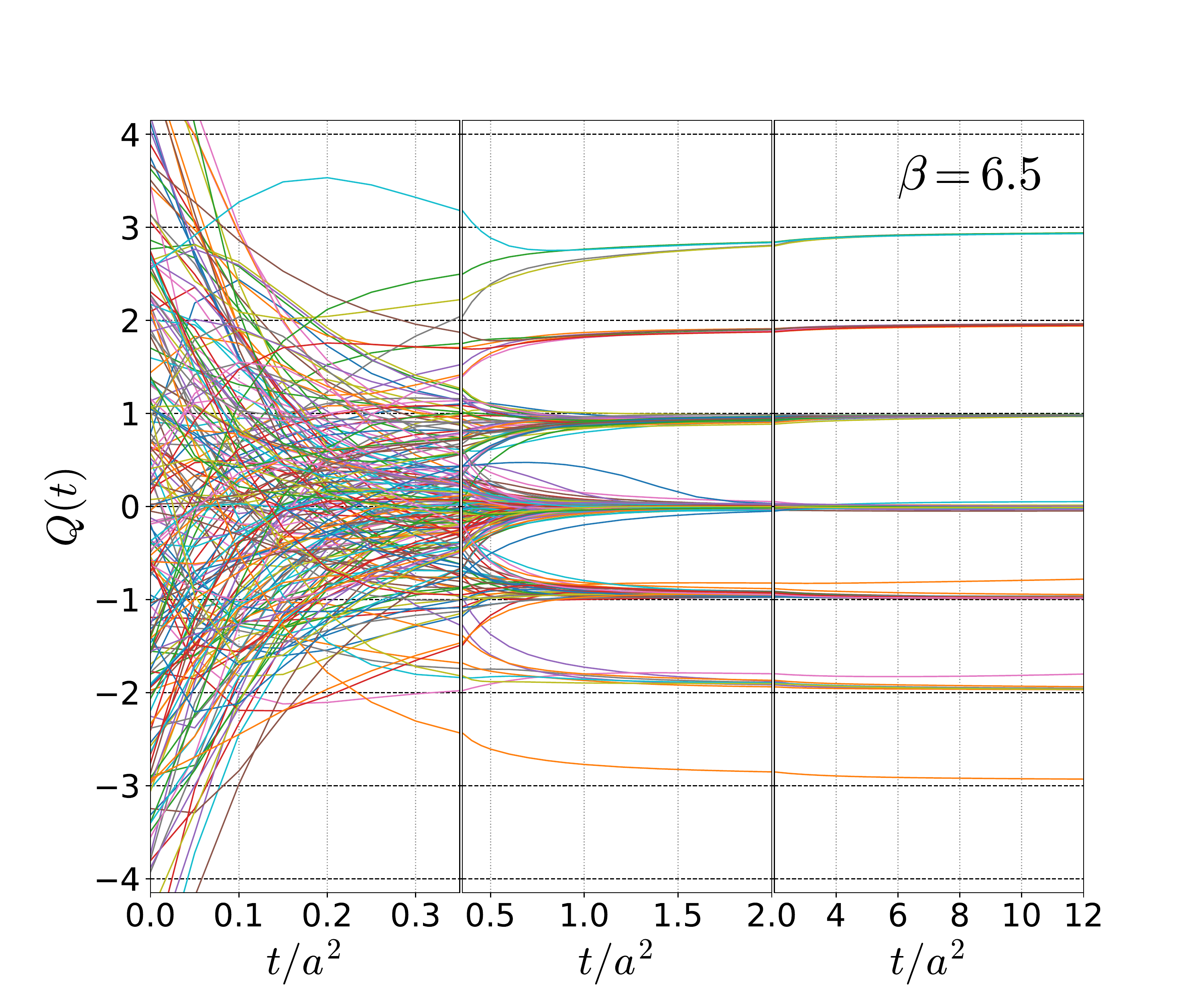}
  \caption{
    Flow time $t$ dependence of $Q(t)$ on $200$ gauge configurations
    at $\beta=6.2$ (left) and $6.5$ (right).
    The range of the flow time $t$ is divided into three panels
    representing $0\le t/a^2 \le0.35$, $0.35\le t/a^2 \le2$,
    and $2\le t/a^2 \le 12$.
  }
  \label{fig:Q(t)}
\end{figure}

\begin{figure}[t]
  \centering
  \vspace{5mm}
  \includegraphics[width=0.45\textwidth,clip]{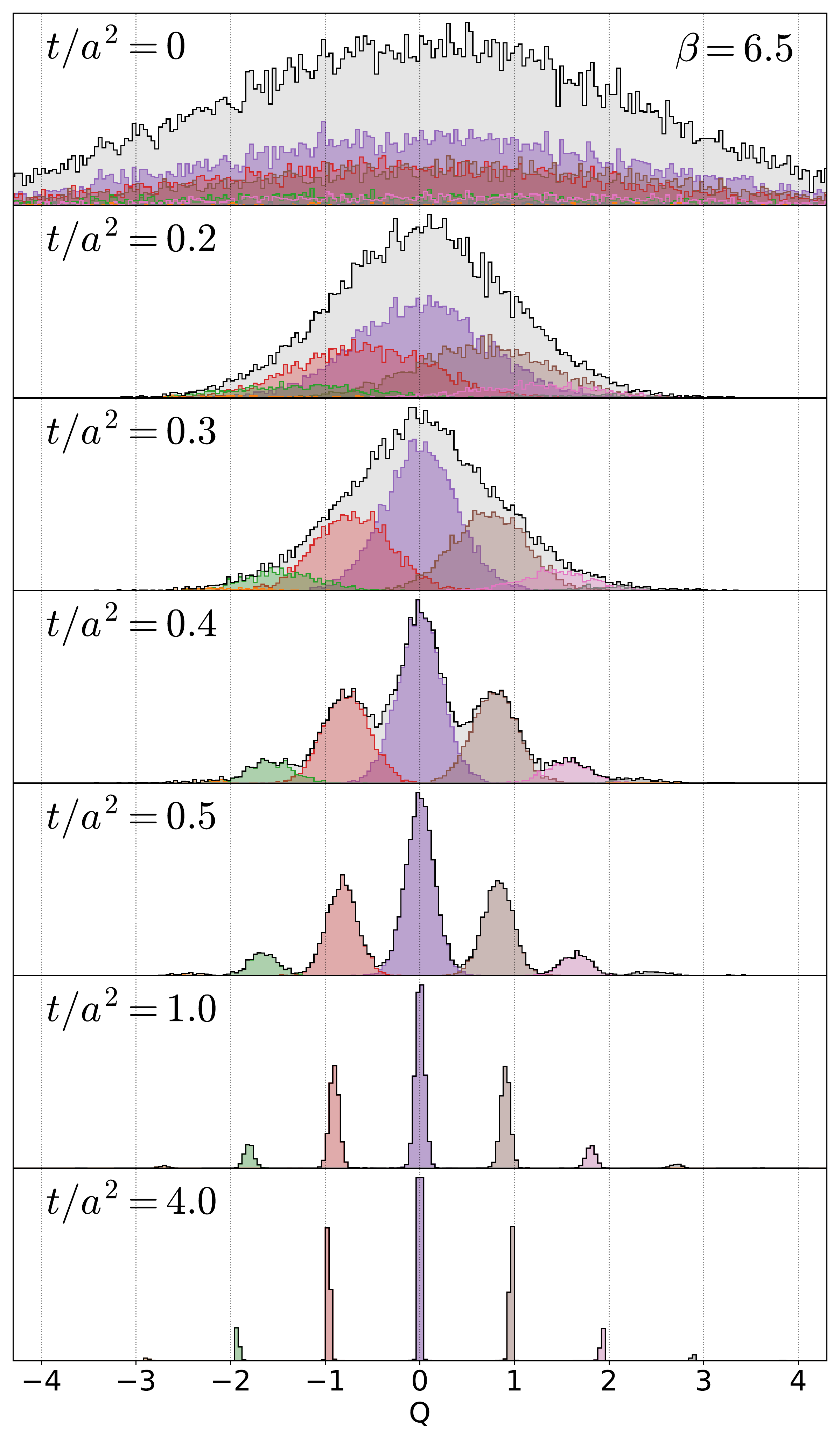}
  \caption{
    Distribution of $Q(t)$ at several values of $t/a^2$.
    The colored histograms are the distributions in individual
    topological sectors; see text.
  }
  \label{fig:Qhist}
\end{figure}

In the present study, we use the gradient
flow~\cite{Luscher:2010iy,Luscher:2011bx} for the smoothing.
The gradient flow is a continuous transformation of the gauge field
characterized by a parameter $t$ called the flow time having
dimension of mass inverse squared.
The gauge field at a flow time $t$ is a smoothed
field with the mean-square smoothing radius
$\sqrt{8t}$~\cite{Luscher:2010iy}.
In the following, we denote 
the topological charge density obtained at $t$ as 
$q_t(x)$, and its four-dimensional integral as
\begin{align}
  Q(t) = \int_V d^4 x  \, q_t(x). 
\end{align}

Shown in Fig.~\ref{fig:Q(t)} is the $t$ dependence of $Q(t)$
calculated on $200$ gauge configurations at 
$\beta=6.2$ and $6.5$.
The horizontal axis shows the dimensionless flow time $t/a^2$
with the lattice spacing $a$.
One finds that the values of $Q(t)$ approach discrete integer values
as $t$ becomes larger.
In Fig.~\ref{fig:Qhist}, we show the distribution of $Q(t)$
for several values of $t/a^2$ by the histogram at $\beta=6.5$.
At $t=0$, the values of $Q(t)$ are distributed continuously around the origin.
As $t$ becomes larger, the distribution converges on discretized
integer values.
For $t/a^2>1.0$, the distribution is almost completely separated
around integer values.
In this range of $t$, 
one can classify the gauge configurations into different topological
sectors labeled by the integer topological charge ${\cal Q}$ defined,
for example, by the nearest integer to $Q(t)$.
It is known that the value of ${\cal Q}$ defined in this way
approximately agrees with the topological charge obtained through
other definitions, and the agreement is better on finer
lattices~\cite{Alexandrou:2017hqw}.

From Figs.~\ref{fig:Q(t)} and \ref{fig:Qhist}, one finds that the
distribution of $Q(t)$ deviates from integer values toward the origin.
This deviation becomes smaller as $t$ becomes larger.
From Fig.~\ref{fig:Q(t)}, one also finds that $Q(t)$ on some
gauge configurations has a ``flipping'' between different topological
sectors;
after $Q(t)$ shows a convergence to an integer value,
it sometimes jumps into another integer~\cite{Alexandrou:2017hqw}.
As this behavior decreases on the finer lattice, 
the flipping would be regarded as a lattice artifact arising from
the ambiguity of the topological sectors on the discretized spacetime.

\begin{table*}
   \centering
   \begin{tabular}{c|ccccccccccc|c}
     \hline
     \hline
     ${\cal Q}$ & -5 & -4 & -3 & -2 & -1 & 0 & 1 & 2 & 3 & 4 & 5
     & $\langle {\cal Q}^2 \rangle$
     \\
     \hline
     $\beta=6.2$ & 2 & 17 & 235 & 1325 & 4571 & 7474 & 4766 & 1352 & 240 & 18 & 0
     & 1.247(15)
     \\
     $\beta=6.5$ & 0 & 5 & 105 & 1080 & 4639 & 8296 & 4621 & 1039 & 202 & 13 & 0
     & 1.039(47)
     \\
     \hline
   \end{tabular}

 \caption{
   Number of the gauge configurations classified into each topological sector
   with the definition of ${\cal Q}$ in Eq.~(\ref{eq:Qdef}).
   The far right column shows the variance of the distribution of ${\cal Q}$.
 }
 \label{table:Qdist}
\end{table*}

In the following, 
  for the numerical definition of the topological charge ${\cal Q}$
  of each gauge configuration we employ
\begin{align}
  {\rm round}[Q(t)]_{t/a^2=4.0},
  \label{eq:Qdef}
\end{align}
where ${\rm round}(x)$ means rounding off to the nearest integer.
As indicated from Fig.~\ref{fig:Q(t)}, the value of ${\cal Q}$ 
hardly changes with the variation of $t/a^2$ in the range $4<t/a^2<12$.
In Table~\ref{table:Qdist}, we show the number of gauge configurations
classified into each topological sector through this definition.
The variance of this distribution $\langle {\cal Q}^2 \rangle$
is shown in the far right column.
In Fig.~\ref{fig:Qhist}, the distributions of $Q(t)$ in
individual topological sectors are shown
by the colored histograms.

\section{Benchmark models}
\label{sec:bench}

In this study, we analyze $q_t(x)$ or $Q(t)$ at small values of $t$
by the ML technique.
Here, $t$ used for the input has to be chosen small enough
that a simple estimate of ${\cal Q}$ like Eq.~(\ref{eq:Qdef})
is not possible.
In this section, before the main analysis with the ML technique
we discuss the accuracy obtained only from $Q(t)$ without ML.
These analyses serve as benchmarks for evaluating the genuine
benefit of ML.

Throughout this study, as the performance metric of a model for
an estimate of ${\cal Q}$ we use the accuracy defined by 
\begin{align}
  P = \frac{\rm number~of~correct~answers}
  {\rm number~of~total~data}.
  \label{eq:P}
\end{align}
Because the numbers of gauge configurations
on different topological sectors differ significantly
as in Table~\ref{table:Qdist}, Eq.~(\ref{eq:P}) would not
necessarily be a good performance metric.
In particular, the topological sector with ${\cal Q}=0$ has 
the largest number, and a model which estimates ${\cal Q}=0$ for all
configurations obtains the accuracy
$P\simeq0.37$ $(0.41)$ for $\beta=6.2$ $(6.5)$,
although such a model is, of course, meaningless.
One has to keep in mind this possible problem of Eq.~(\ref{eq:P}).
In Secs.~\ref{sec:0} and \ref{sec:q(x)},
we use the recalls of individual topological
sectors $R_{\cal Q}$ defined by
\begin{align}
  R_{\cal Q} = \frac{N_{\cal Q}^{\rm correct}}{N_{\cal Q}} ,
  \label{eq:R}
\end{align}
complementary to Eq.~(\ref{eq:P}) to inspect the bias 
of NN models,
where $N_{\cal Q}$ is the number of configurations in the
topological sector ${\cal Q}$ and
$N_{\cal Q}^{\rm correct}$ is the number of correct answers among them.

\begin{figure}[t]
  \centering
  \vspace{5mm}
  \includegraphics[width=0.49\textwidth,clip]{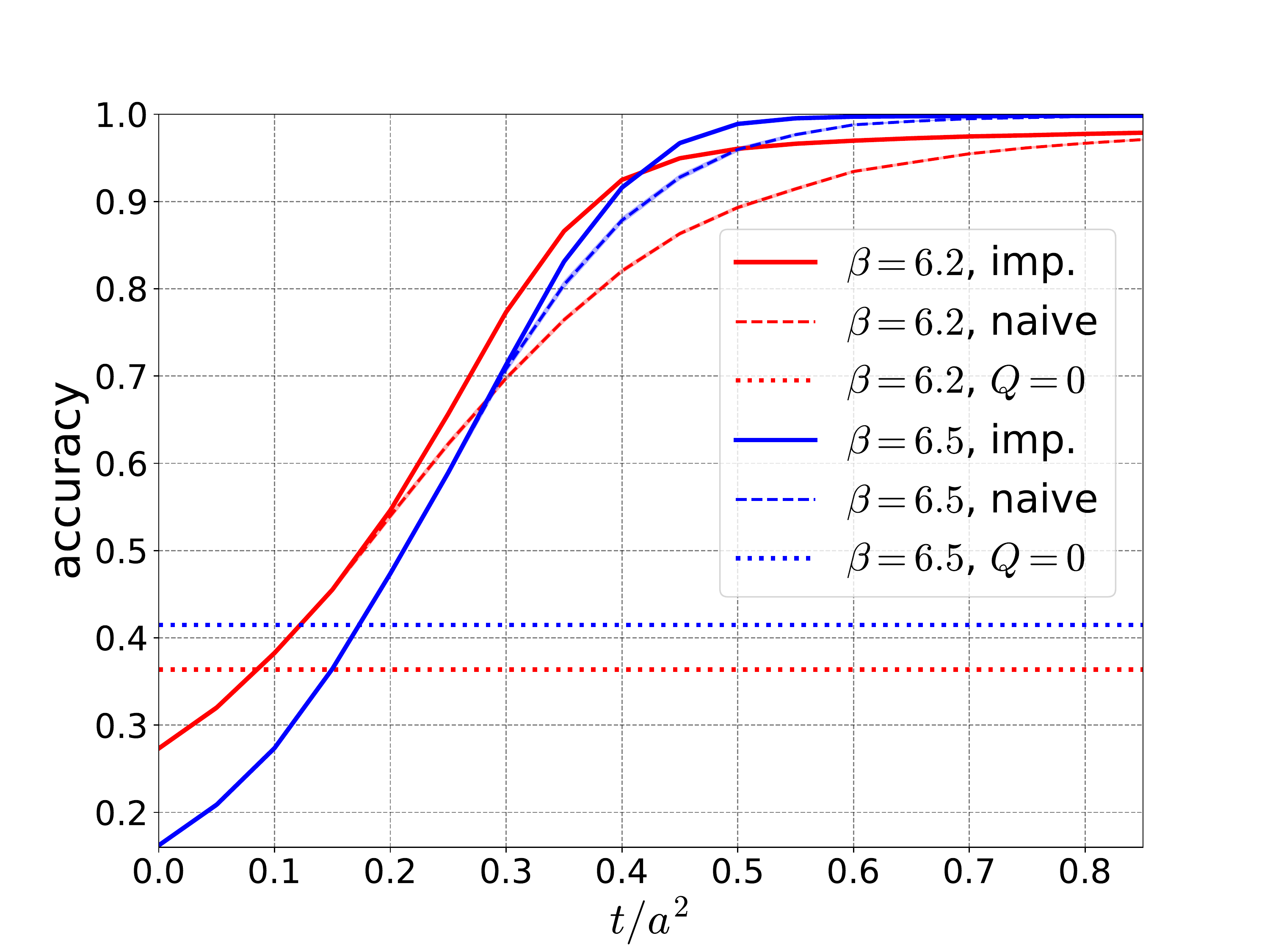}
  \caption{
    Flow time $t$ dependence of the accuracies $P_{\rm naive}$ and
    $P_{\rm imp}$ obtained by the models
    Eqs.~(\ref{eq:Qnaive}) and (\ref{eq:Qimp}), respectively.
    The dotted lines show the accuracy of the model that answers
    ${\cal Q}=0$ for all configurations.
  }
  \label{fig:bench}
\end{figure}

\begin{figure}[t]
  \centering
  \vspace{5mm}
  \includegraphics[width=0.49\textwidth,clip]{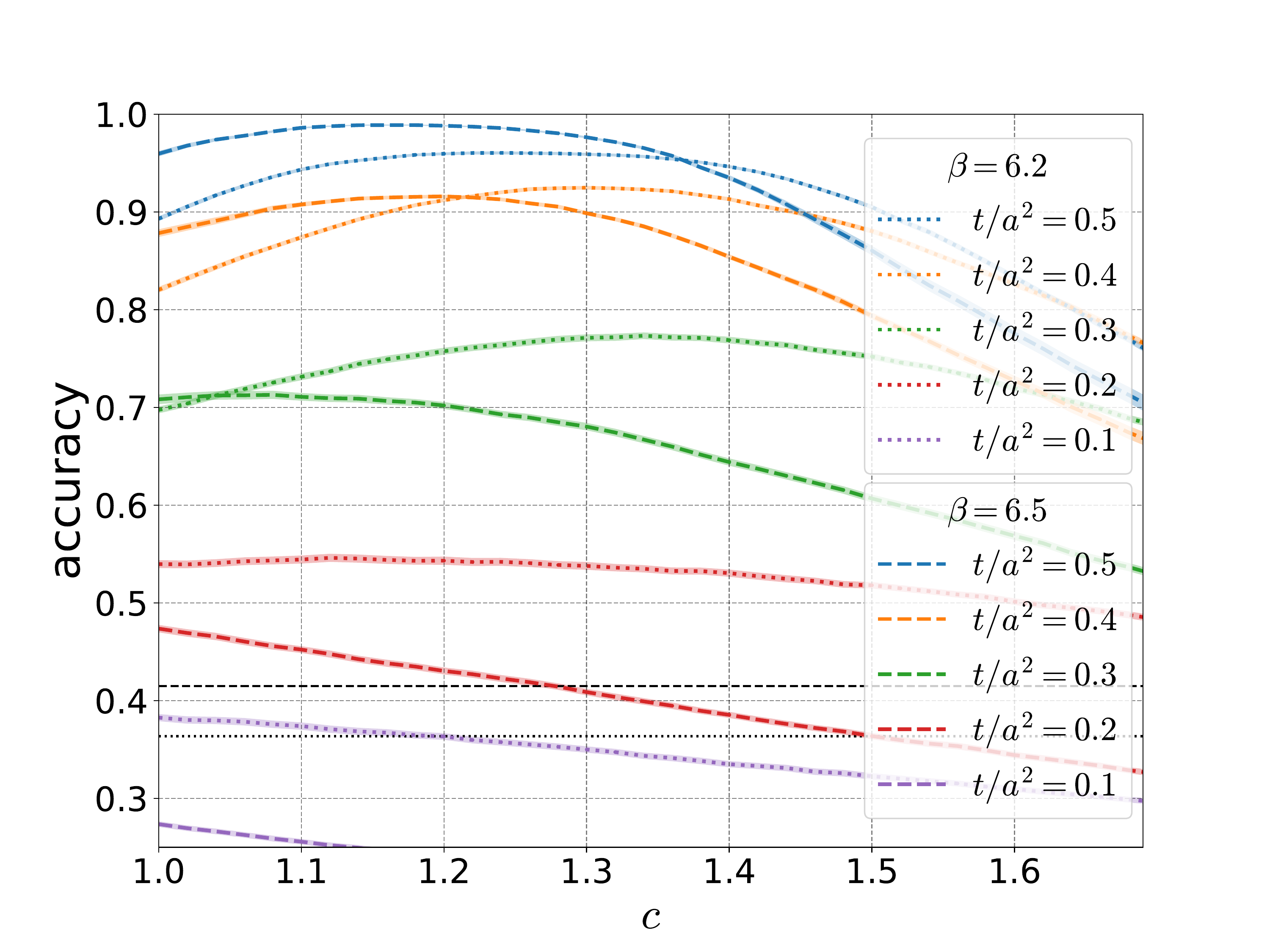}
  \caption{
    Accuracy of Eq.~(\ref{eq:Qimp}) as a function of $c$
    for several values of $t/a^2$.
    The statistical errors are shown by the shaded band, although
    the width of the bands are almost the same as the thickness of the lines.
  }
  \label{fig:c}
\end{figure}

\begin{table}
   \centering
   \begin{tabular}{c|cc}
     \hline
     \hline
     $t/a^2$ & $\beta=6.2$ & $\beta=6.5$   \\
     \hline
     0   & 0.273(3) & 0.162(3) \\
     0.1 & 0.383(4) & 0.274(3) \\
     0.2 & 0.546(4) & 0.474(4) \\
     0.3 & 0.773(3) & 0.713(4) \\
     0.4 & 0.925(2) & 0.916(2) \\
     0.5 & 0.960(1) & 0.989(1) \\
     1.0 & 0.982(1) & 0.999(0) \\
     2.0 & 0.992(1) & 0.999(0) \\
     4.0 & 1.000(0) & 1.000(0) \\
     10.0& 0.993(1) & 0.999(0) \\
     \hline
     \hline
   \end{tabular}

 \caption{
   Accuracy $P_{\rm imp}$ obtained by the model Eq.~(\ref{eq:Qimp})
   with the optimization of $c$.
 }
 \label{table:bench}
\end{table}

To make an estimate of ${\cal Q}$ from $Q(t)$,
we consider two simple models.
The first model is just rounding off $Q(t)$ as
\begin{align}
  Q_{\rm naive}={\rm round} [Q(t)].
  \label{eq:Qnaive}
\end{align}
The accuracy obtained by this model, $P_{\rm naive}$, 
as a function of $t$ is shown in Fig.~\ref{fig:bench} by the dashed lines.
The figure shows that the accuracy of Eq.~(\ref{eq:Qnaive})
approaches $100\%$ as $t/a^2$ becomes larger corresponding to the
behavior of $Q(t)$ in Figs.~\ref{fig:Q(t)} and \ref{fig:Qhist}.
At $t/a^2=4.0$, Eq.~(\ref{eq:Qnaive}) is equivalent to Eq.~(\ref{eq:Qdef}) 
and the accuracy becomes $100\%$ by definition.

The model in Eq.~(\ref{eq:Qnaive}) can be improved
with a simple modification.
In Fig.~\ref{fig:Qhist}, one sees that 
the distribution in each topological sector is shifted
toward the origin from ${\cal Q}$.
This behavior suggests that Eq.~(\ref{eq:Qnaive}) can be
improved by applying a constant before rounding off,
\begin{align}
  Q_{\rm imp}={\rm round}[ cQ(t)] , 
  \label{eq:Qimp}
\end{align}
where $c$ is a parameter determined so as to maximize the accuracy
in the range $c>1$ for each $t$.
In Fig.~\ref{fig:c}, we show the $c$ dependence of the accuracy 
of Eq.~(\ref{eq:Qimp}) for several values of $t/a^2$.
The figure shows that the accuracy has a maximum at $c>1$
for some $t/a^2$.
In this case, the model Eq.~(\ref{eq:Qimp}) has a better accuracy
than Eq.~(\ref{eq:Qnaive}) by tuning the parameter $c$.
We denote the optimal accuracy of Eq.~(\ref{eq:Qimp})
as $P_{\rm imp}$.
In Fig.~\ref{fig:bench}, the $t/a^2$ dependence of $P_{\rm imp}$
is shown by the solid lines.
The numerical values of $P_{\rm imp}$ are depicted in Table~\ref{table:bench}
for some $t/a^2$.
Figure ~\ref{fig:bench} shows that 
a clear improvement of the accuracy by the single-parameter tuning
is observed at $t/a^2\gtrsim 0.2$ and $0.3$ for $\beta=6.2$ and $6.5$,
respectively.
We note that $P_{\rm naive}=P_{\rm imp}=1$ at $t/a^2=4.0$ by definition.
Table~\ref{table:bench} also shows that $P_{\rm imp}$ is almost unity
at $t/a^2=2.0$ and $10.0$, which shows that 
the value of ${\cal Q}$ defined by 
Eq.~(\ref{eq:Qdef}) hardly changes with the variation of $t/a^2$ 
in the range $t/a^2\gtrsim2.0$.

As $P_{\rm imp}$ is already close to unity at $t/a^2=0.5$,
it is difficult to obtain a non-trivial gain in the accuracy
from the analysis of $q_t(x)$ by NN for $t/a^2\ge0.5$.
In the following, therefore, we feed the data at $t/a^2<0.5$
to the NN.
In the following sections, we use $P_{\rm imp}$ for
a benchmark of the accuracy obtained by the NN models.

\section{Learning $Q(t)$}
\label{sec:0}

From this section we employ the ML technique
for the analysis of the lattice data.
As discussed in Secs.~\ref{sec:intro} and \ref{sec:organization},
among various analyses we found that
the most efficient result is obtained when
a set of the values of $Q(t)$ at several $t$
is analyzed by FNN.
In this section, we discuss this result.
The analysis of the multi-dimensional data by CNN 
will be reported in later sections.

\subsection{Setting}
\label{sec:setting0}

\begin{table}
   \centering
   \begin{tabular}{c|cc}
     \hline
     \hline
     layer & output size & activation   \\
     \hline
     input        & 3 & - \\
     full connect & 5 & logistic \\
     full connect & 1 & - \\
     \hline
     \hline
   \end{tabular}
   \caption{
     Design of the FNN used for the analysis of $Q(t)$. 
   }
   \label{table:network0}
\end{table}

In this section we employ a simple FNN model without convolutional
layers. The FNN accepts three values of $Q(t)$ at different $t$
as inputs, and is trained to predict ${\cal Q}$ by 
supervised learning.
The structure of the FNN is shown in Table~\ref{table:network0}.
The FNN has only one hidden layer with five units
that are fully connected with the input and output layers.
We use the logistic (sigmoid) function for 
the activation function of the hidden layer.
Although we have also tried the rectified linear unit (ReLU)
for the activation function,
we found that the logistic function gives a better result.
We employ the regression model, i.e. the output of the FNN
is given by a single real number.
The final prediction of ${\cal Q}$ is then obtained
by rounding off the output to the nearest integer.

For the supervised learning, we randomly divide 20,000 gauge
configurations into 10,000 and two 5,000 sub-groups.
We use 10,000 data for the training, and one of the 5,000 data sets
for the validation analysis. 
The last 5,000 data are used for the evaluation of the accuracy
of the trained NN.
The supervised learning is repeated 10 times with different divisions
of the configurations,
and the uncertainty of the accuracy is estimated from the variance%
\footnote{
    This analysis is called the shuffle-split cross validation.
    There are alternative evaluations of the stability,
    such as the $k$-fold cross validation.
    As the resulting error would not be changed so much,
    however, we only use the shuffle-split cross validation
    throughout this study.
}.

We use the mean-squared error for the loss function, and minimize it 
through the updates of the NN parameters 
by the ADAM~\cite{2014arXiv1412.6980K} with the default setting.
The update is repeated for 3,000 epochs with batchsize 16.
The optimized parameter set of the FNN is then determined as the one
giving the lowest value of the loss function on the validation data.

The FNN is implemented by the Chainer framework~\cite{Chainer}.
The training of the FNN in this section has been carried out 
as a single-core job on a XEON processor (Xeon E5-2698-v3).
It takes about 40 minutes for a single training
in this environment.

\subsection{Result}
\label{sec:Q(t)result}

\begin{table}
   \centering
   \begin{tabular}{l|cc}
     \hline
     \hline
     input $t/a^2$ & $\beta=6.2$ & $\beta=6.5$   \\
     \hline
     0.45, 0.4, 0.35  & 0.974(2) & 0.998(1) \\
     0.4, 0.35, 0.3   & 0.975(2) & 0.997(1) \\
     0.35, 0.3, 0.25  & 0.967(2) & 0.996(1) \\
     {\bf 0.3, 0.25, 0.2}   & {\bf 0.959(2)} & {\bf 0.990(2)} \\
     0.25, 0.2, 0.15  & 0.939(3) & 0.951(2) \\
     0.2, 0.15, 0.1   & 0.864(3) & 0.831(5) \\
     0.15, 0.1, 0.05  & 0.692(4) & 0.647(8) \\
     0.1, 0.05, 0     & 0.538(5) & 0.499(6) \\
     \hline
     0.4, 0.3, 0.2    & 0.971(2) & 0.995(1) \\
     0.3, 0.2, 0.1    & 0.941(2) & 0.957(2) \\
     0.2, 0.1, 0      & 0.741(3) & 0.682(4) \\
     \hline
     \hline
   \end{tabular}
   \caption{
     Accuracy of the trained FNN in Table~\ref{table:network0}
     with various sets of the input data.
     Left column shows the values of $t/a^2$
     that evaluate $Q(t)$ for the input.
     Errors are estimated from the variance among 10 different trainings.
   }
   \label{table:result0}
\end{table}

Shown in Table~\ref{table:result0} are the accuracies 
obtained by the trained FNN for various choices of the input data.
The left column shows the set of three flow times $t/a^2$
that evaluate $Q(t)$ used for the input of the FNN.
In the upper eight rows we show the results
with the input flow times
$t/a^2=(\hat{t}_{\rm max}, \hat{t}_{\rm max}-0.05, \hat{t}_{\rm max}-0.1)$.
The table shows that the accuracy is improved as $\hat{t}_{\rm max}$
becomes larger.
By comparing this result with Table~\ref{table:bench}
one finds that the accuracy obtained by the FNN is significantly
higher than $P_{\rm imp}$ at $t/a^2=\hat{t}_{\rm max}$.
In particular, the accuracy at $\hat{t}_{\rm max}=0.3$, i.e. 
$t/a^2=(0.3,0.25,0.2)$, shown in bold is 
as high as $99\%$ for $\beta=6.5$, while the benchmark model
Eq.~(\ref{eq:Qimp}) gives $P_{\rm imp}\simeq0.71$.
This result shows that the prediction of ${\cal Q}$ 
from the numerical data at $t/a^2\le0.3$
is remarkably improved with the aid of the ML technique.
Table~\ref{table:result0} also shows that 
the accuracy improves further as $\hat{t}_{\rm max}$ becomes larger,
but the improvement from $P_{\rm imp}$ is limited for much larger
$\hat{t}_{\rm max}$ because $P_{\rm imp}$ is already close to unity.
The same tendency is obtained for $\beta=6.2$, although the accuracy
is slightly lower than $\beta=6.5$.

\begin{figure}[t]
  \centering
  \vspace{5mm}
  \includegraphics[width=0.49\textwidth,clip]{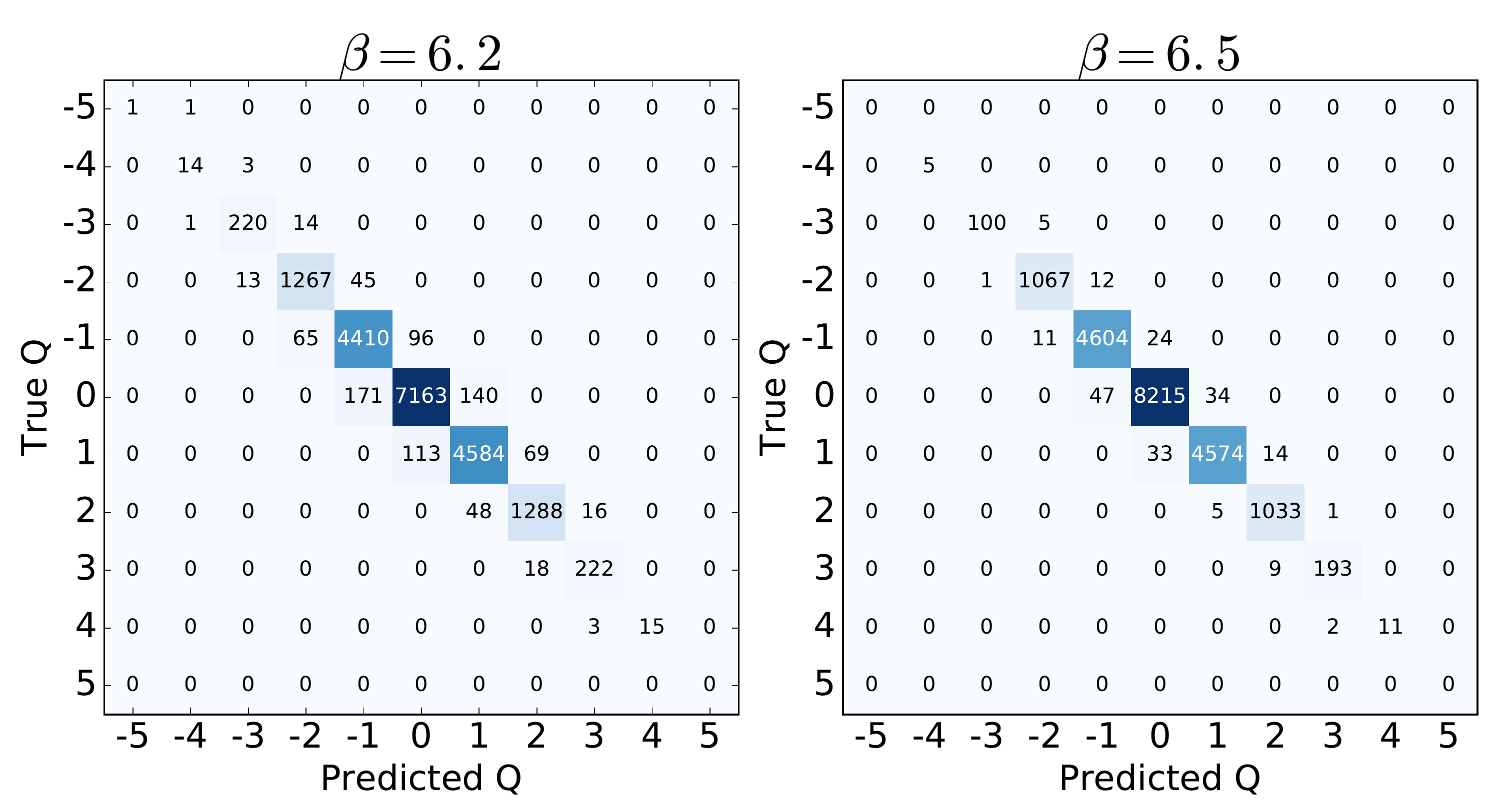}
  \caption{
    Confusion matrix of the trained FNN model with the input $Q(t)$ at
    $t/a^2=(0.3,0.25,0.2)$ for $\beta=6.2$ (left) and $6.5$ (right).
    Each cell shows the number of configurations with the true and
    predicted values of ${\cal Q}$.
  }
  \label{fig:Qscatter}
\end{figure}
\begin{table*}
   \centering
   \begin{tabular}{c|c|ccccccccc}
     \hline
     \hline
     & & \multicolumn{9}{c}{$R_{\cal Q}$}
     \\
     \cline{3-11}
     $\beta$ & $P$
     & -4 & -3 & -2 & -1 & 0 & 1 & 2 & 3 & 4 
     \\
     \hline
     6.2 & 0.959
     & 0.912 & 0.933 & 0.953 & 0.964 & 0.959 & 0.964 & 0.947 & 0.919 & 0.949 
     \\
     6.5 & 0.990
     & 1.000 & 0.951 & 0.987 & 0.993 & 0.991 & 0.988 & 0.992 & 0.946 & 0.818 
     \\
     \hline
     \hline
   \end{tabular}
   \vspace{5mm}
\caption{
   Accuracy $P$ and the recalls of individual topological
   sectors $R_{\cal Q}$ obtained by the FNN with the input $Q(t)$ at
   $t/a^2=(0.3,0.25,0.2)$.
 }
 \label{table:R_Q}
\end{table*}

In Fig.~\ref{fig:Qscatter} we show the confusion matrix
that plots the numbers of configurations with the true and predicted
values of ${\cal Q}$ with the input flow times $t/a^2=(0.3,0.25,0.2)$
for $\beta=6.2$ and $6.5$.
From the figure one finds that the deviation of the output of
the FNN from the true value is at most $\pm1$.
In Table~\ref{table:R_Q}, the recalls from Eq.~(\ref{eq:R}) of individual
topological sectors are shown. The table shows that $R_{\cal Q}$ tends
to be suppressed as $|{\cal Q}|$ becomes larger, but its ${\cal Q}$ 
dependence is mild. This result suggests that the training of
the FNN is successful in spite of the biased ${\cal Q}$ distribution
of the training data. See also the analysis in Sec.~\ref{sec:Q<=3}.

The high accuracy obtained in this analysis is not surprising qualitatively.
As shown in Fig.~\ref{fig:Q(t)}, for many configurations the behavior
of $Q(t)$ is monotonic at $0.2\le t/a^2\le0.3$.
Therefore, the value at large $t$ can be estimated easily, 
for example, by the human eye, for almost all configurations.
It is reasonable to interpret that the FNN learns this behavior.
We, however, remark that the accuracy of $99\%$ obtained by the trained FNN
is still non-trivial.
We have tried to find a simple function to predict ${\cal Q}$ from
three values of $Q(t)$ at $t/a^2=(0.3, 0.25, 0.2)$, and
also performed blind tests with human beings.
These trials have been able to obtain $95\%$ accuracy easily,
but have failed in attaining $99\%$; see Appendix~\ref{sec:app}
for more discussion on this point.
These results suggest that ML finds non-trivial features
in the behavior of $Q(t)$.

In the lower three rows of Table~\ref{table:result0},
we show the accuracies of the trained FNN with the input flow times 
$t/a^2=(\hat{t}_{\rm max}, \hat{t}_{\rm max}-0.1,\hat{t}_{\rm max}-0.2)$
for several values of $\hat{t}_{\rm max}$.
These accuracies are slightly lower than the results with
$t/a^2=(\hat{t}_{\rm max}, \hat{t}_{\rm max}-0.05, \hat{t}_{\rm max}-0.1)$
at the same $\hat{t}_{\rm max}$.
We have also tested the FNN models analyzing four values of $Q(t)$.
It, however, was found that the accuracy does not exceed 
the case with three input data with the same maximum $t/a^2$.
We have also tested FNN models with more complex structure,
for example, with multiple hidden layers.
A statistically significant improvement of the accuracy,
however, was not observed here either.

In the conventional analysis of ${\cal Q}$ with the gradient flow
discussed in Sec.~\ref{sec:Q},
one uses the value of $Q(t)$ at a large flow time
at which the distribution of $Q(t)$ is well localized.
This means that the gradient flow equation has to be solved
numerically~\cite{Luscher:2010iy} up to the large flow time.
Moreover, concerning the continuum $a\to0$ limit 
it is suggested that the flow time has to be fixed in physical
units~\cite{Giusti:2018cmp}, which means that 
the flow time in lattice units, $t/a^2$, becomes large
as the continuum limit is approached.
On the other hand, our analysis with the aid of the FNN can estimate
${\cal Q}$ successfully only with the data at $t/a^2\lesssim0.3$,
and thus the numerical cost for solving the gradient flow 
can be reduced\footnote{
  In Yang-Mills theories this cost reduction is practically beneficial.
  The most numerically demanding part in typical Yang-Mills lattice simulations
  except for the gradient flow is the gauge updates, because the cost
  for the measurement of gauge observables is usually negligible.
  The numerical costs of the gauge updates and the gradient flow are 
  dominated by the calculation of staples.
  From the numbers to calculate staples in both algorithms
  one finds that the cost for the gradient flow up to $t/a^2\simeq4.0$
  corresponds to the gauge updates of ${\cal O}(100)$ sweeps.
  In Yang-Mills theories measurements of gauge observables are usually
  performed with much shorter Monte-Carlo separations,
  $\lesssim{\cal O}(10)$, because
  it is responsible for the noise reduction.
  This is true even when the auto-correlation length of ${\cal Q}$ is
  large, because statistical fluctuation of gauge observables are typically
  significantly larger than the ${\cal Q}$ dependence of the observables.
  To realize such frequent measurements, the cost reduction
  for the measurement of ${\cal Q}$ from ${\cal O}(100)$ sweeps
  is highly desirable.
}.

Table~\ref{table:result0} shows that better accuracy is obtained 
on the finer lattice (larger $\beta$).
From Fig.~\ref{fig:Q(t)}, it is suggested that this tendency 
comes from the reduction of the ``flipping'' of $Q(t)$ on the finer lattice,
as the non-monotonic flipping makes the prediction of ${\cal Q}$
from $Q(t)$ at small $t/a^2$ difficult.
This effect is also suggested from Fig.~\ref{fig:bench}, as $P_{\rm naive}$
and $P_{\rm imp}$ at $\beta=6.2$ are lower than those at $\beta=6.5$.
We note that this lattice spacing dependence hardly changes even if
we scale the value of $t$ to determine ${\cal Q}$ in Eq.~(\ref{eq:Qdef})
in physical units if $t$ is sufficiently large.
Provided that the flipping of $Q(t)$ comes from the lattice artifact
related to the ambiguity of the topological sectors on the discretized
spacetime, it is conjectured that the imperfect accuracy of
the FNN is to a large extent attributed to this lattice artifact.
Then, the imperfect accuracy of the FNN at finite $a$ is inevitable,
and the accuracy should become better as the lattice spacing
becomes finer.
Therefore, it is conjectured that the systematic uncertainty arising from
the imperfect accuracy of the FNN is suppressed in the analysis of
the continuum extrapolation.

\subsection{Susceptibility and higher-order cumulants}

Next, we consider the variance of the topological charge,
$\langle {\cal Q}^2 \rangle$, which is related to the 
topological susceptibility as $\chi_Q=\langle {\cal Q}^2 \rangle/V$.
From the output of the FNN with the input flow times $t/a^2=(0.3,0.25,0.2)$,
the variance of ${\cal Q}$ is calculated to be
\begin{align}
  &\langle {\cal Q}^2 \rangle_{\rm NN} = 1.253(15)(2) \quad({\rm for}~~\beta=6.2),
  \label{eq:<Q^2>NN62}\\
  &\langle {\cal Q}^2 \rangle_{\rm NN} = 1.037(46)(1) \quad({\rm for}~~\beta=6.5),
  \label{eq:<Q^2>NN65}
\end{align}
where the first and second errors represent the statistical error
obtained by the jackknife analysis 
and the uncertainty of the FNN model estimated from 
$10$ different trainings, respectively.
These values agree well with those shown in Table~\ref{table:Qdist}.

From the output of the FNN
the fourth- and sixth-order cumulants of ${\cal Q}$ are calculated to be
$\langle {\cal Q}^4 \rangle_{c,\rm NN} = 0.30(6)(0)$ and 
$\langle {\cal Q}^6 \rangle_{c,\rm NN} = -1.16(41)(12)$ for $\beta=6.2$,
and 
$\langle {\cal Q}^4 \rangle_{c,\rm NN} = 0.35(11)(1)$ and 
$\langle {\cal Q}^6 \rangle_{c,\rm NN} = -0.82(46)(1)$ for $\beta=6.5$.
On the other hand, the cumulants extracted from the distribution
in Table~\ref{table:Qdist} are
$\langle {\cal Q}^4 \rangle_c = 0.38(7)$ and 
$\langle {\cal Q}^6 \rangle_c = -1.11(49)$ for $\beta=6.2$,
and 
$\langle {\cal Q}^4 \rangle_c = 0.39(12)$ and 
$\langle {\cal Q}^6 \rangle_c = -0.82(52)$ for $\beta=6.5$.
One thus finds that the values of
$\langle {\cal Q}^4 \rangle_{\rm c}$ and $\langle {\cal Q}^6 \rangle_{\rm c}$
estimated by the FNN are consistent with the original values 
within the statistics.
These results, of course, do not mean that the FNN can make a 
prediction for arbitrary higher-order cumulants.
In fact, Fig.~\ref{fig:Qscatter} suggests that the accuracy
becomes worse for large $|{\cal Q}|$.
This behavior would modify higher-order cumulants more strongly.
We also note that these results are obtained only with specific
setups of the lattice simulation. There are no guarantees of
obtaining a similar stability when the setup,
for example, the lattice volume and gauge action, is changed.
It, however, is worth emphasizing that 
the FNN developed in this study can reproduce even the sixth-order
cumulant within the statistics for at least two simulation setups.

\subsection{Training with restricted data sets}
\label{sec:Q<=3}

\begin{figure}[t]
  \centering
  \vspace{5mm}
  \includegraphics[width=0.49\textwidth,clip]{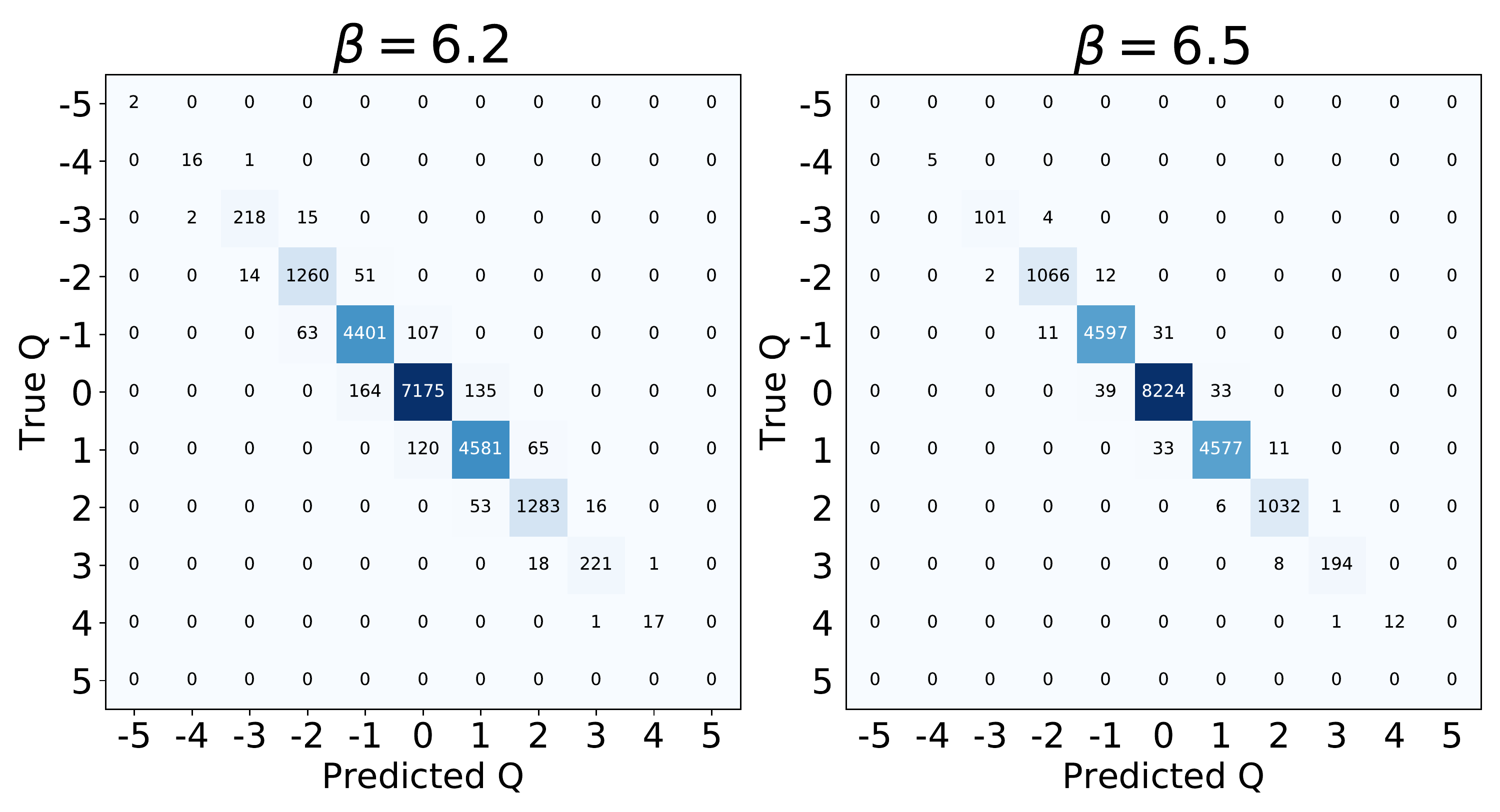}
  \caption{
    Confusion matrix of the FNN model trained only with the training data
    at $|{\cal Q}|\le3$.
    The input flow times are $t/a^2=(0.3,0.25,0.2)$.
  }
  \label{fig:Qscatter2}
\end{figure}

As shown in Table~\ref{table:Qdist},
the probability of the appearance of large $|{\cal Q}|$ is severely suppressed.
Therefore, it is practically possible that the
FNN encounters a large $|{\cal Q}|$ that the FNN has never experienced
during the training phase.
In this subsection, we study the response of the FNN for 
${\cal Q}$ that is not included in the training data.

For this purpose, we prepare training and validation data
sets composed only of $|{\cal Q}|\le3$.
We then perform the supervised learning
of the FNN using these restricted data sets.
The input flow times are $t/a^2=(0.3,0.25,0.2)$.
The setup of the supervised learning is the same as before
except for the restriction of the training data.

In Fig.~\ref{fig:Qscatter2}, we show examples of the confusion matrix of the
FNN models trained by the data with $|{\cal Q}|\le3$ for $\beta=6.2$ and $6.5$.
The total accuracy of the FNN is $P=0.959$ and $0.990$ for
$\beta=6.2$ and $6.5$, respectively, which are consistent with
the training without the restriction.
From Fig.~\ref{fig:Qscatter2} one finds that the trained FNN can predict
a correct answer even for large $|Q|$ that are not included in the
training data.
Moreover, comparison with Fig.~\ref{fig:Qscatter}
shows that the accuracies for the topological sectors 
$|{\cal Q}|=4$ and $5$ hardly change with and without the restriction.

This result is not surprising.
First, we employ a regression model for the FNN, i.e. the output
of the FNN is given by a single number, rather than a classification
model that outputs probabilities of predetermined categories.
Therefore, the answers of the FNN are not restricted to specific values.
Second, from the behaviors of $Q(t)$ in Fig.~\ref{fig:Q(t)}
this result is also expected as $Q(t)$ in different 
topological sectors behaves differently around $0.2\le t/a^2 \le 0.3$.
Finally, as in Table~\ref{table:Qdist} 
the number of configurations with $|{\cal Q}|=4$ and $5$ are small
in the data set;
$0.18\%$ and $0.09\%$ for $\beta=6.2$ and $6.5$, respectively.
Their effects on the supervised learning thus should be small
even if they are included in the training data.

We, however, stress that this result
does not mean that the trained FNN can make a prediction 
for arbitrarily large $|{\cal Q}|$ successfully.
As the result in this subsection is obtained only for particular data sets,
in general one has to be careful when applying NN for the analysis of
the data that are not included in the training data.

\subsection{Reduction of training data}
\label{sec:Ntrain}

\begin{table}
   \centering
   \begin{tabular}{c|ccccc}
     \hline
     \hline
     $N_{\rm train}$ & $10,000$ & $5,000$ & $1,000$ & $500$ & $100$  \\
     \hline
     $\beta=6.2$  & $0.959(2)$ & $0.959(2)$ & $0.959(2)$ & $0.953(3)$ & $0.903(7)$ \\
     $\beta=6.5$  & $0.990(2)$ & $0.990(2)$ & $0.989(2)$ & $0.989(1)$ & $0.902(8)$ \\
     \hline
     \hline
   \end{tabular}
   \caption{
     Dependence of the accuracy of the trained FNN
     on the number of the training data $N_{\rm train}$.
   }
   \label{table:reduction}
\end{table}

So far, we have performed the training of the FNN with 
the number of training data points $N_{\rm train}=10,000$.
Now we consider training with much smaller $N_{\rm train}$.

Shown in Table~\ref{table:reduction} is the accuracy of the
trained FNN with various $N_{\rm train}$ with
the input flow times $t/a^2=(0.3,0.25,0.2)$.
The structure of the FNN is the same as before.
From the table, one finds that the FNN is successfully trained
even with $N_{\rm train}=500$.

This result shows that the cost for
the preparation of the training data can be reduced.
The reduction of $N_{\rm train}$ is also responsible for reducing 
the numerical cost for the training.
With $N_{\rm train}=10,000$, the training of the FNN requires
about $40$~minutes on a single core of a XEON processor,
while only $5.5$~minutes is needed with $N_{\rm train}=500$
on the same environment.

\subsection{Robustness}
\label{sec:robust}

\begin{table}
   \centering
   \begin{tabular}{|c|c|c|c|}
     \hline
     \multicolumn{2}{|c|}{} & \multicolumn{2}{c|}{ analyzed data } \\
     \cline{3-4}
     \multicolumn{2}{|c|}{} & $\beta=6.2$ & $\beta=6.5$ \\
     \hline
     \multirow{3}{*}{ \shortstack{training \\ data} }
     & $\beta=6.2$ & $0.959(2)$ & $0.986(2)$ \\
     \cline{2-4}
     & $\beta=6.5$ & $0.956(2)$ & $0.990(2)$ \\
     \cline{2-4}
     & $6.2/6.5$ & $0.958(1)$ & $0.989(2)$ \\
     \hline

   \end{tabular}
   \caption{
     Accuracy obtained by the analysis of the data with
     different $\beta$ from the one used for the training.
     Last row shows the accuracy of the FNN trained by
     the mixed data set at $\beta=6.2$ and $6.5$.
   }
   \label{table:robustness}
\end{table}

Next, we consider an application of the FNN for the 
analysis of data obtained at different $\beta$
from the one used for the training.
In Table~\ref{table:robustness}, we show the accuracy obtained
with various combinations of the $\beta$ values used for training
and analysis with the input flow times $t/a^2=(0.3,0.25,0.2)$.
The table shows that the accuracy becomes worse when the different
data set is analyzed, but the reduction is small and almost
within the statistics.
We have also performed the training of the FNN with a combined
data set of $\beta=6.2$ and $6.5$. The result of this analysis is
shown in the far bottom row in Table~\ref{table:robustness}.
One finds that this FNN can predict ${\cal Q}$ for each $\beta$
with the same accuracy within the statistics as those trained
for individual $\beta$.

From these results it is conjectured that a NN trained by 
the combined data sets at $\beta=6.2$ and $6.5$ can analyze
the gauge configurations in the parameter range $6.2\le\beta\le6.5$
obtained by the Wilson gauge action with a high accuracy.
Moreover, the range of $\beta$ would be extended by performing
the supervised learning with the data at different $\beta$ values.
Once such a model is developed, it can be used for the analysis of
various $\beta$ values and will play a useful role.

The two lattices studied here
have almost the same spatial volume in physical units.
Therefore, in the present study the robustness against 
variation of the spatial volume in physical units is not studied.
To check the volume dependence, one has to generate gauge configurations
with different spatial volumes and repeat the same analysis.
This analysis is left for future work.

Because the analysis in the present study is performed only for SU(3)
gauge theory with the Wilson gauge action, 
it is not clear from this analysis if 
the FNN gains high accuracy for other lattice simulations,
for example, at nonzero temperature, with different lattice gauge
action, and QCD with dynamical fermions.
However, the results of other lattice
simulations suggest that the qualitative behavior of $Q(t)$ is similar to 
Figs.~\ref{fig:Q(t)} and \ref{fig:Qhist}~\cite{Ce:2015qha,
  Taniguchi:2016tjc,Alexandrou:2017hqw} even if the setting of the
simulation is changed.
It thus is na\"ively expected that our method can be used 
successfully even for different lattice simulations, although
in this case the FNN has to be trained using the data obtained
by the specific simulation.
Although in Sec.~\ref{sec:Q(t)result} we found that the input flow
times $t/a^2=(0.3,0.25,0.2)$ are an optimal choice for the case
of SU(3) gauge theory with the Wilson gauge action, the optimal
set of input flow times will also be different for different
simulations.

\section{Learning topological charge density $q_t(x)$}
\label{sec:q(x)}

In this section we employ a CNN and train it to analyze 
the four-dimensional field $q_t(x)$.
Motivation for this analysis is the search for characteristic
features responsible for the topology 
in the four-dimensional space by ML.
As the CNN has been applied to image recognition quite
successfully~\cite{Krizhevsky:2012,Szegedy_2015_CVPR,Simonyan15,He_2016_CVPR},
it is expected that the features in four-dimensional
space can also be captured by this network, and an improvement
of the accuracy compared with the previous section would be observed.
In particular, if the quantum gauge configurations have 
local structures like instantons~\cite{Weinberg:1996kr},
such structures would be recognized by the CNN and 
used for an efficient prediction of ${\cal Q}$.

\subsection{Input data}
\label{sec:input4}

Let us first discuss the choice of input data for the CNN.
The gauge configurations on the lattice are
described by the link variables $U_\mu(x)$, which are elements of
the group SU(3).
The most fundamental choice for the input data is the link variables.
However, as $U_\mu(x)$ is described by $72$ real variables per lattice site,
reduction of the data size is desirable for efficient training.
Moreover, because physical observables are given only by gauge-invariant
combinations of $U_\mu(x)$, 
the CNN must learn the concept of gauge invariance,
and accordingly SU(3) matrix algebra, so that it can 
make a successful prediction of ${\cal Q}$ from $U_\mu(x)$.
These concepts, however, would be too complicated for simple CNN models.

In the present study, for these reasons we use the topological charge
density $q_t(x)$ as the input to the CNN.
$q_t(x)$ is gauge invariant, and 
the degrees of freedom per lattice site is one.
To suppress the size of the input data further, 
we reduce the lattice volume to $8^4$ from $16^4$ and $24^4$
by average pooling as preprocessing
assuming that the typical spatial size of the features in the
four-dimensional space is large enough and they are not
spoiled by this pooling procedure.
In addition to the analysis of $q_t(x)$ at a single $t$ value,
we prepare a combined data set of $q_t(x)$ with several values of $t$
and analyze it as multi-channel data with the CNN.

\subsection{Designing CNN}
\label{sec:network4}

\begin{table}
   \centering
   \begin{tabular}{c|ccc}
     \hline
     \hline
     layer & filter size & output size & activation   \\
     \hline
     input       & -     & $8^d\times N_{\rm ch}$ & - \\
     convolution & $3^d$ & $8^d\times 5$ & logistic \\
     convolution & $3^d$ & $8^d\times 5$ & logistic \\
     convolution & $3^d$ & $8^d\times 5$ & logistic \\
     global average pooling & $8^d$ & $1\times5$ & - \\
     full connect & - & 5 & logistic \\
     full connect & - & 1 & - \\
     \hline
     \hline
   \end{tabular}
   \caption{
     Design of the CNN for the analysis of the multi-dimensional data.
     The dimension $d$ of the input data is $4$ in Sec.~\ref{sec:q(x)}.
     In Sec.~\ref{sec:reduction}, we analyze the data with 
     $d=1,2,3$ obtained by the dimensional reduction.
   }
   \label{table:network4}
\end{table}

In this section, we use a CNN 
with convolutional layers that deal with four-dimensional data.
In Table~\ref{table:network4}, we show the structure of the CNN model,
where $d$ denotes the dimension of the spacetime 
and is set to $d=4$ throughout this section.
The model has three convolutional layers with filter size $3^4$ and
five output channels.
In the convolutional layers, we use periodic padding
for all directions to respect the periodic boundary conditions
of the gauge configuration.
$N_{\rm ch}$ denotes the number of channels of input data per lattice site;
$N_{\rm ch}=1$ when $q_t(x)$ at a single $t$ is fed into the CNN.
We also perform multi-channel analysis by feeding $q_t(x)$ 
at $N_{\rm ch}$ flow times.

Lattice gauge theory has translational symmetry, 
and a shift of the spatial coordinates 
toward any direction does not change the value of ${\cal Q}$.
To ensure that the CNN automatically respects this property,
we insert a global average pooling (GAP) layer~\cite{2013arXiv1312.4400L}
after the convolutional layers.
The GAP layer takes the average with respect to the spatial coordinates
for each channel.
The output of the GAP layer is then processed by two fully-connected
layers before the final output.
The logistic activation function is used for the convolutional and
fully-connected layers.

The training of the CNN in this section has been mainly carried out 
on Google Colaboratory~\cite{Colaboratory}.
We use 12,000 data for the training, 2,000 data for the validation, and 
6,000 data for the test, respectively.
The batchsize for the minibatch training is 200.
We repeat the parameter tuning for 500 epochs.
Other settings of the training are the same as 
the previous section.

Besides the CNN model in Table~\ref{table:network4},
we have tested various variations of the model.
For example, we tested the ReLU activation function in place of the logistic one.
The use of the fully-connected layer in place of the GAP layer and
convolutional layers with the $5^4$ filter size is also tried.
The number of output channels of the convolutional layers was varied
up to $20$.
We, however, found that these variations do not improve the accuracy
at all, while they typically increase the numerical cost for
the training.
The CNN in Table~\ref{table:network4} is a simple but efficient
choice among all these variations.

\subsection{Results}
\label{sec:result4}

\begin{table*}
   \centering
   $\beta=6.2$
   \begin{tabular}{c|c|c|ccccccccc}
     \hline
     \hline
     & & & \multicolumn{9}{c}{$R_{\cal Q}$}
     \\
     \cline{4-12}
     $N_{\rm ch}$ & input $t/a^2$ & $P$
     & -4 & -3 & -2 & -1 & 0 & 1 & 2 & 3 & 4 
     \\
     \hline
     1 & 0 & 0.371    & 0 & 0 & 0 & 0 & 1.000 & 0 & 0 & 0 & 0
     \\
     1 & 0.1 & 0.401  & 0 & 0 & 0.002 & 0.255 & 0.702 & 0.341 & 0.008 & 0 & 0 
     \\
     1 & 0.2 & 0.552  & 0 & 0.043 & 0.240 & 0.495 & 0.687 & 0.597 & 0.336 & 0.111 & 0
     \\
     1 & 0.3 & 0.776  & 0 & 0.391 & 0.687 & 0.760 & 0.821 & 0.794 & 0.740 & 0.569 & 0 
     \\
     3 & 0.3,0.2,0.1 & 0.942
     & 0.200 & 0.913 & 0.944 & 0.950 & 0.944 & 0.939 & 0.937 & 0.889 & 0.571
     \\
     \hline
     \hline
   \end{tabular}
   \vspace{5mm}
   
   $\beta=6.5$
   \begin{tabular}{c|c|c|ccccccccc}
     \hline
     \hline
     & & & \multicolumn{9}{c}{$R_{\cal Q}$}
     \\
     \cline{4-12}
     $N_{\rm ch}$ & input $t/a^2$ & $P$
     & -4 & -3 & -2 & -1 & 0 & 1 & 2 & 3 & 4 
     \\
     \hline
     1 & 0 & 0.388    & 0 & 0 & 0 & 0 & 1.000 & 0 & 0 & 0 & 0
     \\
     1 & 0.1 & 0.396  & 0 & 0 & 0 & 0.086 & 0.889 & 0.129 & 0 & 0 & 0
     \\
     1 & 0.2 & 0.479  & 0 & 0 & 0.108 & 0.445 & 0.641 & 0.459 & 0.150 & 0 & 0
     \\
     1 & 0.3 & 0.698  & 0 & 0.170 & 0.585 & 0.730 & 0.727 & 0.701 & 0.624 & 0.395 & 0.071  
     \\
     3 & 0.3,0.2,0.1 & 0.953
     & 0 & 0.830 & 0.951 & 0.956 & 0.952 & 0.962 & 0.968 & 0.953 & 0.286
     \\
     \hline
     \hline
   \end{tabular}

 \caption{
   Accuracy $P$ and the recalls of individual topological
   sectors $R_{\cal Q}$ obtained by the analysis of the topological charge density
   in the four-dimensional space by the CNN.
   The input data has $N_{\rm ch}$ channels.
 }
 \label{table:4single}
\end{table*}

In Table~\ref{table:4single}, we show the performance of the
trained CNN with various inputs.
Left two columns show $N_{\rm ch}$ and the flow time(s) used for the
input.
On the upper four rows, the results with $N_{\rm ch}=1$
with the input data at $t/a^2=0$, $0.1$, $0.2$, and $0.3$ are shown.
The last row shows the result of the multi-channel analysis with
$N_{\rm ch}=3$ where $q_t(x)$ at $t/a^2=(0.3, 0.2, 0.1)$ are used.
The third column shows the accuracy $P$ of the trained CNN
obtained for each input.
In the table, we also show 
the recalls of individual topological sectors $R_{\cal Q}$
defined in Eq.~(\ref{eq:R}).

The top row of Table~\ref{table:4single} shows $P$ and $R_{\cal Q}$
obtained by the analysis of the topological charge density of the
original gauge configuration without the gradient flow.
Although we obtain a nonzero $P$,
the recall of each ${\cal Q}$ shows that in this case
the CNN is trained to answer ${\cal Q}=0$ for almost all configurations.
This means that the CNN fails in obtaining any features
responsible for the determination of ${\cal Q}$.

Next, the results with $N_{\rm ch}=1$ but nonzero $t/a^2$ show
that $P$ becomes larger with increasing $t/a^2$.
From $R_{\cal Q}$ one also finds that 
the output of the CNN scatters on different topological sectors.
However, by comparing $P$ with that of the benchmark model
$P_{\rm imp}$ in Table~\ref{table:bench} with the same $t/a^2$,
one finds that $P$ and $P_{\rm imp}$ are almost the same.
This result suggests that the CNN is trained to answer
$Q_{\rm imp}$ and no further information is obtained from
the analysis of the four-dimensional data of $q_t(x)$.

Finally, from the multi-channel analysis with the input flow times
$t/a^2=(0.3, 0.2, 0.1)$,
one finds that the accuracy $P$ is significantly
enhanced from the case with $N_{\rm ch}=1$ and exceeds $94\%$ for each $\beta$.
However, this accuracy is the same within the error as that obtained
by the FNN in Sec.~\ref{sec:0} with $t/a^2=(0.3, 0.2, 0.1)$
shown in Table~\ref{table:result0}.
This result implies that the CNN is trained to
obtain $Q(t)$ for each $t$ and then predicts the answer from them 
with a similar procedure as the FNN in Sec.~\ref{sec:0}.

From these results, it is conjectured that
our analyses of four-dimensional data by CNN 
failed in finding structures in the four-dimensional space
responsible for the determination of ${\cal Q}$.
The numerical cost for the training of the CNN in this section is
a few orders larger than in Sec.~\ref{sec:0}, although 
a clear improvement of the accuracy is not observed.
Therefore, for practical purposes
the analysis in the previous section with the FNN is superior.
We also note that this negative result on the analysis of 
four-dimensional data would be improved by changing the design of
the CNN.
For example, as the down-sampling of the data by the average pooling
would smear out small features, direct analysis of the original
data without the average pooling would modify the result.
We leave this analysis for future study.

\section{Dimensional reduction}
\label{sec:reduction}

\begin{figure}[t]
  \centering
  \vspace{5mm}
  \includegraphics[width=0.49\textwidth,clip]{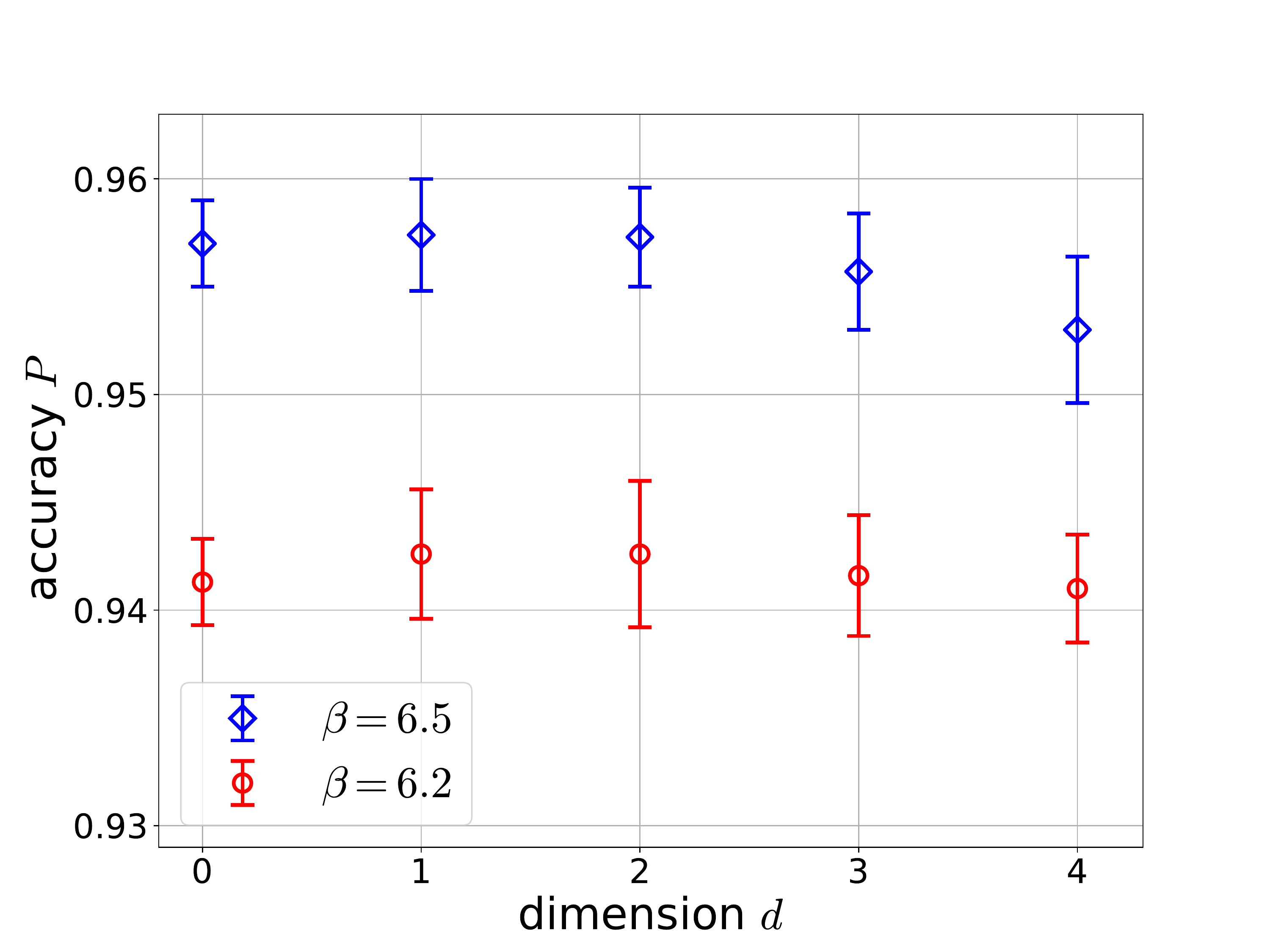}
  \caption{
    Dependence of the accuracy $P$ on the spacetime dimension $d$
    after the dimensional reduction.
    The values at $d=0$ and $4$ corresponds to the results
    in Secs.~\ref{sec:0} and \ref{sec:q(x)}, respectively.
  }
  \label{fig:dim_red}
\end{figure}

In the previous two sections we discussed the analysis of
the four-dimensional topological charge density $q_t(x)$ and
its four-dimensional integral $Q(t)$ by ML.
The spatial dimensions of these input data are $d=4$ and $0$,
respectively.
In this section, we analyze data with dimensions $d=1$--$3$
obtained by dimensional reduction by the CNN.

We consider the integral of the topological charge density
with respect to some coordinates
\begin{align}
  &\tilde{q}_t^{(3)}(x_1,x_2,x_3) = \int dx_4 \, q_t(x_1,x_2,x_3,x_4) ,
  \\
  &\tilde{q}_t^{(2)}(x_1,x_2) = \int dx_4 dx_3 \, q_t(x_1,x_2,x_3,x_4) ,
  \\
  &\tilde{q}_t^{(1)}(x_1) = \int dx_4 dx_3 dx_2 \, q_t(x_1,x_2,x_3,x_4) ,
\end{align}
with $q_t(x)= q_t(x_0,x_1,x_2,x_3)$.
Here, $\tilde{q}_t^{(d)}$ is the $d$-dimensional field 
analyzed by the CNN.
The structure of the CNN is the same as in the previous section
(see Table~\ref{table:network4}) except for the value of $d$.
The procedure of the supervised learning is also the same.
We analyze the multi-channel data with $N_{\rm ch}=3$
and $t/a^2=(0.3,0.2,0.1)$.

In Fig.~\ref{fig:dim_red}, we show 
the accuracy obtained by the analysis of the $d$-dimensional data
$\tilde{q}_t^{(d)}$ by CNN.
The data points at $d=0$ show the result obtained by the analysis
of $Q(t)$ by FNN in Sec.~\ref{sec:0}
with $t/a^2=(0.3,0.2,0.1)$ given in Table~\ref{table:result0}.
From the figure, one finds that
the accuracy does not have a statistically significant $d$ dependence,
although the results at $d=1$ and $2$ would be slightly better
than $d=0$.
This result supports our conjecture in the previous section
that the CNN fails in finding characteristic features in the
multi-dimensional data.

\section{Discussion}

In the present study, we have investigated the application of the machine
learning techniques to the classification of the topological sector
of gauge configurations in SU(3) Yang-Mills theory.
The Wilson gauge action has been used for generating gauge configurations.
The topological charge density $q_t(x)$ at zero and nonzero flow
times $t$ is used as input to the neural networks (NN)
with and without dimensional reduction.

We found that the prediction of the topological charge ${\cal Q}$
can be made most efficiently when $Q(t)$ at small flow times is used
as the input of the NN.
In particular, we found that the value of ${\cal Q}$ defined 
at a large flow time can be predicted with high accuracy
only with $Q(t)$ at $t/a^2 \le 0.3$;
at $\beta=6.5$, the accuracy exceeds $99\%$.
This result suggests that 
the numerical cost of solving the gradient flow toward large
flow times would be omitted in the analysis of ${\cal Q}$
with the aid of ML.
It will be an interesting future study to pursue this possibility further
by additional analyses.

Because the prediction of the NN does not have $100\%$ accuracy, 
the analysis of ${\cal Q}$ by NN gives rise to uncontrollable
systematic uncertainties.
However, our analyses indicate that the accuracy is improved 
as the continuum limit is approached.
Moreover, as discussed in Sec.~\ref{sec:0}, the imperfect accuracy
would to a large extent come from intrinsic uncertainty of the
topological sectors on the lattice with finite $a$.
It thus is expected that the analysis of ${\cal Q}$ becomes
more accurate as the lattice spacing becomes finer.
As the $99\%$ accuracy is already attained at $\beta=6.5$
($a\simeq0.044$~fm), the analysis with the ML will
be used safely for $\beta\gtrsim6.5$.

All numerical analyses in this study have been carried out
in SU(3) Yang-Mills theory with the Wilson gauge action.
While the dependence on the lattice spacing, $a$, has been studied,
the lattice volume in physical units is fixed.
It thus is not clear from the present study if ML can be 
successfully applied to lattice simulations with different settings, 
such as QCD with dynamical fermions, different lattice gauge action
and different lattice volumes.
As discussed in Sec.~\ref{sec:robust}, however, it is notable that
the behaviors of $Q(t)$ obtained by different lattice
simulations are similar to one another.
It thus is na\"ively expected that our method is applicable 
even for different lattice simulations, although
this conjecture has to be checked explicitly on each setting.

We found that the analysis of the multi-dimensional field 
$q_t(x)$ by CNN can gain high accuracy.
However, an improvement in accuracy
compared with the analysis of $Q(t)$ by FNN was not observed
using the CNN employed in Sec.~\ref{sec:q(x)} and \ref{sec:reduction}.
Moreover, in Sec.~\ref{sec:reduction} it was found that the accuracy
does not have a statistically significant dependence on
the dimension of the input data.
A plausible interpretation of this result is that
the CNN employed in the present study fails
in capturing useful structures in four-dimensional space
relevant for the determination of ${\cal Q}$.
It will be an interesting future work to pursue the recognition of 
structures in four-dimensional space using ML.
One straightforward extension in this direction is 
analysis with a CNN having a more complex structure.
In particular, analysis without preprocessing by average
pooling will improve the performance if this preprocessing smears out
the small features in the multi-dimensional space.
Another interesting direction is analysis of gauge
configurations at high temperatures where the dilute instanton-gas
picture is applicable.
As the topological charge should be carried by well-separated local
objects at such temperatures, the search for the multi-dimensional
space by CNN would be easier than the vacuum configurations.
It is also interesting to analyze $q_t(x)$ at a large flow time
after subtracting the average, because the NN can no longer make
use of the information on $Q(t)$ by such preprocessing.
We leave these analyses for future research.

The authors thank A.~Tomiya for many useful discussions.
They also thank H.~Fukaya and K.~Hashimoto.
The lattice simulations of this study are in part carried out
on OCTOPUS at the Cybermedia Center, Osaka University and
Reedbush-U at Information Technology Center, The University of Tokyo. 
The NN are constructed on the Chainer framework.
The supervised learning of the NN in Secs.~\ref{sec:q(x)} and
\ref{sec:reduction} is in part carried out on Google Colaboratory.
This work was supported by JSPS KAKENHI Grant Numbers~17K05442 and
19H05598.


\appendix

\section{Behavior of $Q(t)$}
\label{sec:app}

\begin{figure}[t]
  \centering
  \vspace{5mm}
  \includegraphics[width=0.49\textwidth,clip]{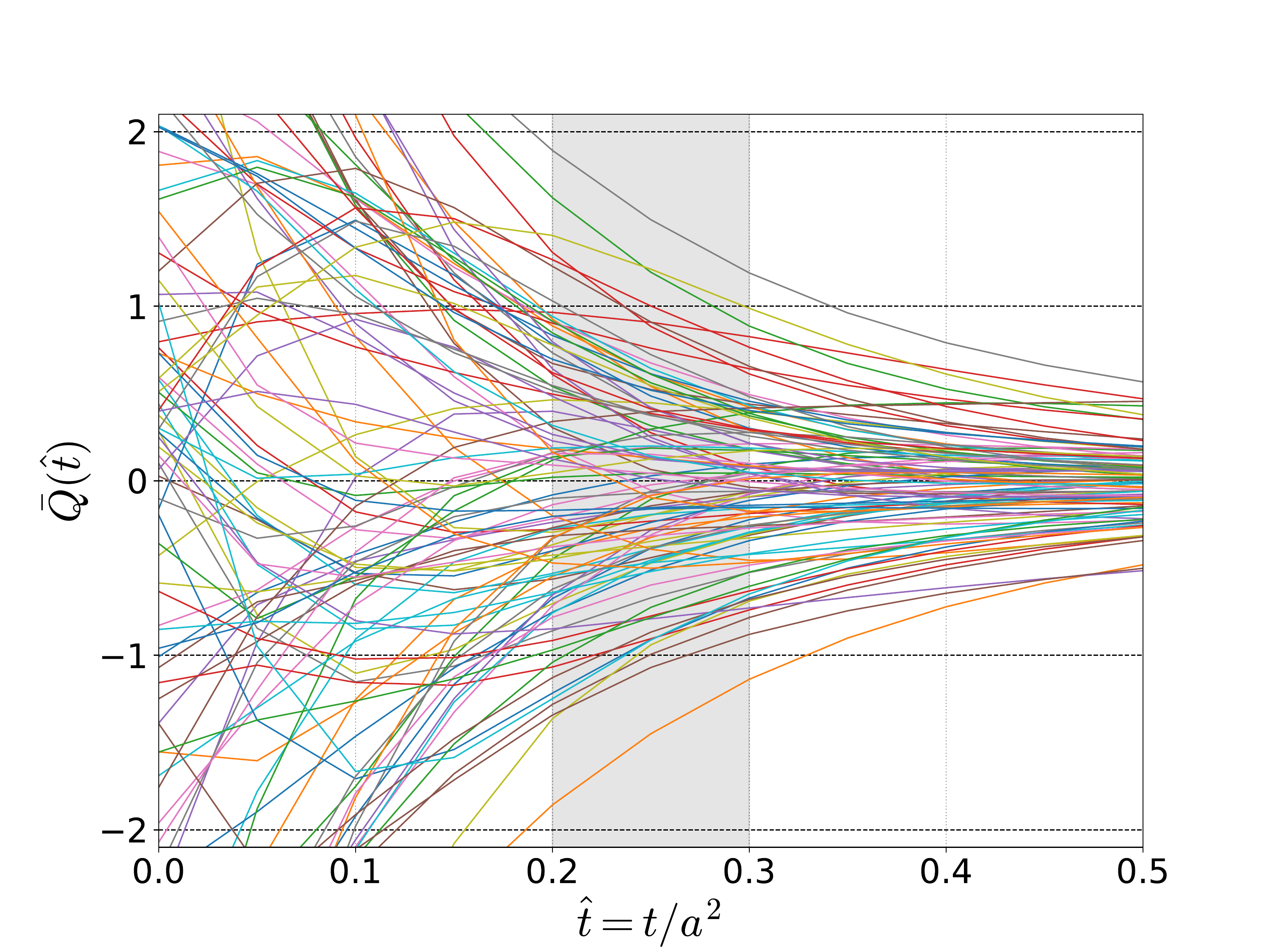}
  \includegraphics[width=0.49\textwidth,clip]{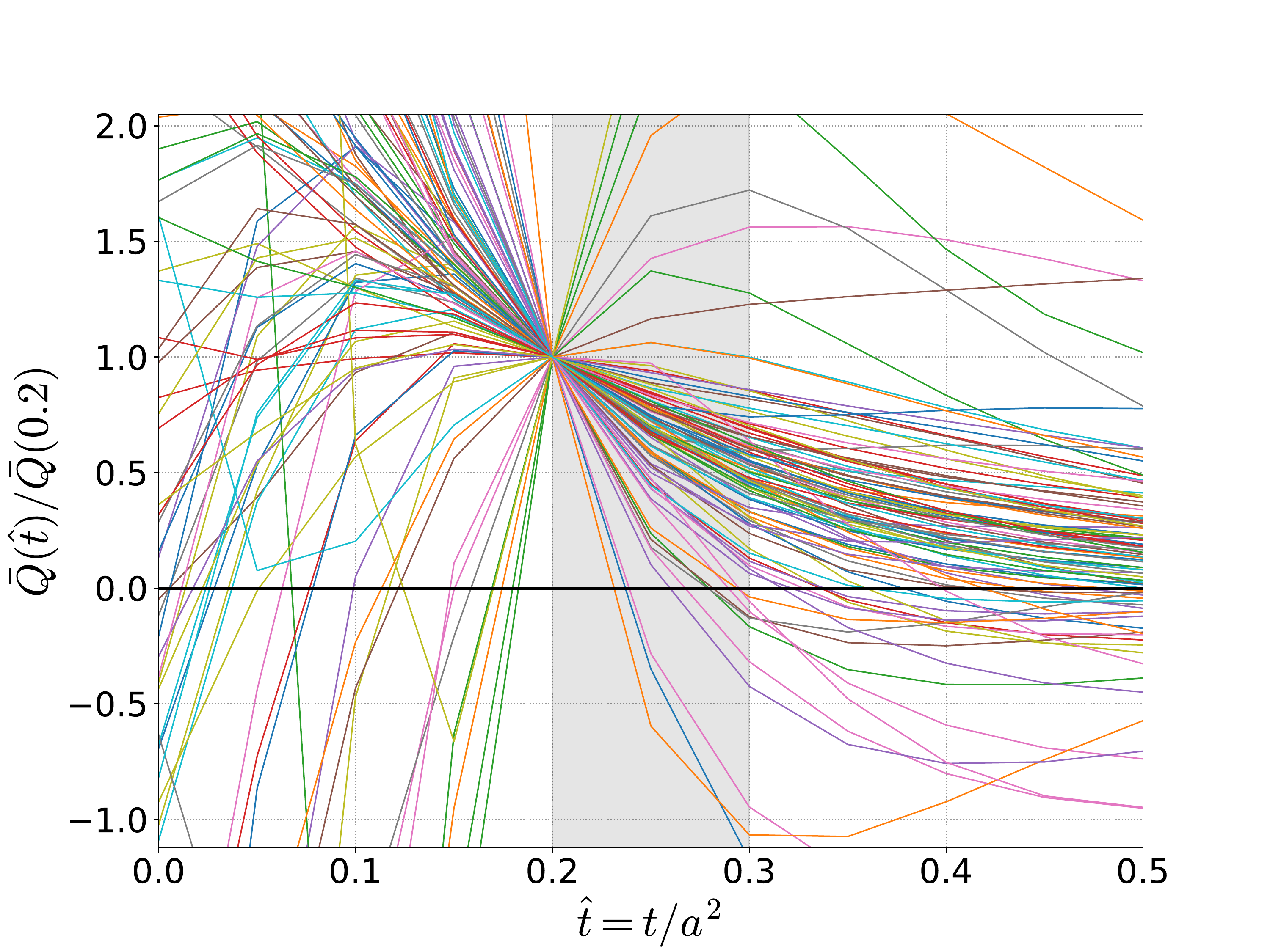}
  \caption{
    Closer look at the behavior of $Q(t)$ at $\hat{t}=t/a^2\le0.5$.
  }
  \label{fig:Q(t)app}
\end{figure}

In this appendix, we take a closer look at the
behavior of $Q(t)$ at small $t$.
In Fig.~\ref{fig:Q(t)app}, we show the $t$ dependence 
of $Q(t)$ on $100$ gauge configurations
at $\beta=6.5$ in two different ways.
In the upper panel we show 
\begin{align}
  \bar{Q}(\hat{t}) = Q(\hat{t}a^2)-{\cal Q},
  \label{eq:barQ}
\end{align}
with $\hat{t}=t/a^2$,
while the lower panel shows
\begin{align}
  \frac{\bar{Q}(\hat{t}) }{\bar{Q}(0.2)}.
  \label{eq:barQ02}
\end{align}
Equation~(\ref{eq:barQ02}) becomes unity at $\hat{t}=0.2$.

In Sec.~\ref{sec:0} it is shown that the trained NN can
estimate the value of ${\cal Q}$ from the behavior of $Q(t)$
at $0.2\le \hat{t} \le0.3$ with $99\%$ accuracy for $\beta=6.5$.
This range of $\hat{t}$ is highlighted by the gray band
in Fig.~\ref{fig:Q(t)app}.
From the upper panel, one sees that $\bar{Q}(\hat{t})$ approaches
zero monotonically on almost all configurations.
However, the panel shows that some lines deviate from this trend.
As a result, it seems difficult to predict the value of ${\cal Q}$
with $99\%$ accuracy (${\cal Q}$ has to be predicted correctly on
99 lines among 100 in the panel) by a simple function 
or the human eye from the behavior at $0.2\le \hat{t}\le0.3$, 
although $95\%$ accuracy is not difficult to attain.
A similar observation is also obtained from the lower panel.
It thus is indicated that the $99\%$ accuracy obtained by the NN 
in Sec.~\ref{sec:Q} is not a trivial result.

\bibliographystyle{ptephy}
\bibliography{topol_ref}

\end{document}